\documentclass[12pt,a4paper,reqno]{article}

\usepackage{amsmath}
\usepackage{amsfonts}
\usepackage{amssymb}
\usepackage{amsthm}
\usepackage{mathbbol}
\usepackage{geometry}
\usepackage{graphicx}
\usepackage{hyperref}
\usepackage[l2tabu, orthodox]{nag}
\usepackage[numbers,square]{natbib}
\usepackage{pgf}
\usepackage{pgfplots}
\usepackage[subrefformat=parens,labelformat=parens]{subfig}
\usepackage{tikz}

\usetikzlibrary{arrows}
\usetikzlibrary{matrix}

\makeatletter
\@ifundefined{chapter}
   {\newtheorem{theorem}{Theorem}[section]}
   {}
\makeatother

\newtheorem*{exercise*}{Exercise}

\AtBeginDocument{
    
    \let\obar\bar

    \let\bar\undefined

    \newcommand{\bar}[1]{\ensuremath{\overline{#1}}}						
}

\def\ba{\begin{align}}
\def\ea{\end{align}}

\newcommand{\mcal}[1]{\ensuremath{\mathcal{#1}}}							
\newcommand{\mrm}[1]{\ensuremath{\mathrm{#1}}}							

\newcommand{\phy}{\ensuremath{\varphi}}									

\newcommand{\norm}[1]{\ensuremath{\left \lVert #1 \right \rVert}}		

\newcommand{\bracketb}[1]{\ensuremath{\big < #1 \big >}}				

\begin{document}

\title{Variational Integrators for Inertial Magnetohydrodynamics}

\author{
\large{Michael Kraus}\\
\small{(michael.kraus@ipp.mpg.de)}
\vspace{.5em}\\
\normalsize{Max-Planck-Institut f\"ur Plasmaphysik}\\
\normalsize{Boltzmannstra\ss{}e 2, 85748 Garching, Deutschland}%
\vspace{.5em}\\
\normalsize{Technische Universit\"at M\"unchen, Zentrum Mathematik}\\
\normalsize{Boltzmannstra\ss{}e 3, 85748 Garching, Deutschland}%
\vspace{1em}\\
}

\date{\today}

\maketitle

\begin{abstract}

Recently, an extended version of magnetohydrodynamics that incorporates electron inertia, dubbed inertial magnetohydrodynamics, has been proposed.
This model features a noncanonical Hamiltonian formulation with a number of conserved quantities, including the total energy and modified versions of magnetic and cross helicity.
In this work, a variational integrator is presented which preserves these conservation laws to machine accuracy.
As long as effects due to finite electron mass are neglected, the scheme preserves the magnetic field line topology so that unphysical reconnection is absent.
Only when effects of finite electron mass are added, magnetic reconnection takes place.
The excellent conservation properties of the method are illustrated by numerical examples in 2D.

\end{abstract}


\newpage

\tableofcontents

\newpage

\section{Introduction}

Ideal magnetohydrodynamics (MHD) is one of the most widely applied theories in laboratory as well as astrophysical plasma physics~\cite{Schnack:2009,GoedbloedPoedts:2004,Biskamp:2003,Davidson:2001,Freidberg:1987}.
Although the system of equations is rather simple, it can used to describe many different macroscopic phenomena like equilibrium states in tokamaks or stellarators, large scale turbulence, or dynamos that generate magnetic fields of stars and planets.
In addition, the system is endowed with a rich geometric structure. It is a Hamiltonian system~\cite{Morrison:1980io}, described in terms of noncanonical Poisson brackets, which have several Casimir invariants associated to them. It has a variational structure, both in Lagrangian~\cite{Newcomb:1962} and Eulerian~\cite{Holm:1998ws} coordinates. But it has also interesting topological properties like the frozen-in magnetic flux~\cite{ArnoldKhesin:1998}.
Even though ideal MHD is applicable to an impressive number of problems, its regime of validity is limited. Therefore various extended MHD models have been derived over the years.
One member of this family of models is inertial MHD~\cite{Lingam:2015}, which adds effects of finite electron inertia to the ideal MHD model, allowing for example for the study of collisionless reconnection. While such an extension was long known for reduced MHD~\cite{Schep:1994}, the extension of the ideal model came along only very recently.
An interesting feature of the inertial MHD model is that it has almost the same Hamiltonian structure as the ideal MHD model, just expressed in terms of a modified magnetic field variable. That is, it has the same kind of Casimir invariants and it is also energy-preserving.

The simplicity of the ideal MHD system combined with this rich geometric structure makes it an ideal prototyping system for the development of structure-preserving numerical algorithms.
By now, several such algorithms for ideal MHD have been proposed.
\citet{LiuWang:2001} approached the problem by coupling the MAC scheme~\cite{Harlow:1965} for the Navier-Stokes equation with Yee's scheme~\cite{Yee:1966} for the Maxwell equations.
\citet{Gawlik:2011} used a discrete Euler-Poincar\'{e} principle (see also the work of~\citet{Pavlov:2011}), which yields a similar scheme as that of \citeauthor{LiuWang:2001}, but with different time discretisation. While this is the most natural discretisation approach  in the Eulerian framework, it is currently not easily possible to obtain higher-order integrators or work in different numerical framework like finite elements or isogeometric analysis.
A variational integrator in Lagrangian variables, based on directly discretising Newcomb's Lagrangian~\cite{Newcomb:1962}, has been derived by \citet{Zhou:2014}. In the Lagrangian framework, this provides a very natural discretisation with excellent conservation properties. Unfortunately, its applicability is somewhat limited, as in many problems the distortion of the mesh will quickly lead to a deterioration of the numerical solution.
Recently, \citet{KrausMaj:2017} proposed a variational discretisation in Eulerian variables based on a formal Lagrangian formulation~\cite{AthertonHomsy:1975, Ibragimov:2006, KrausMaj:2015} combined with ideas from discrete differential forms~\cite{Robidoux:2011, Desbrun:2008, Hirani:2003}. While the discretisation of the variational formulation leads to exact conservation of energy, magnetic helicity and cross helicity, preserving the differential forms character of the physical variables ensures that the divergence of the magnetic field is preserved and prevents checker-boarding, a spurious phenomenon often observed with finite difference discretisations of incompressible fluid equations.
In this paper, the work of~\citet{KrausMaj:2017} is extended towards the inertial MHD model. 

We proceed as follows. In Section~\ref{sec:ideal_mhd}, we start by reviewing the ideal incompressible MHD equations, their Hamiltonian formulation and important conservation laws. We review the concept of formal Lagrangians, which constitutes the starting point for the derivation of variational integrators and show how to apply this concept to the ideal MHD equations. Finally, we describe the modifications of the ideal MHD equations that lead to the inertial MHD system.
In Section~\ref{sec:vi}, we explain the variational discretisation and the staggered grid approach, which is motivated by discrete differential forms. Here, we use the same approach as \citet{KrausMaj:2017}, but we describe the discretisation in a more heuristic way, that should be understandable also without in-depth knowledge of differential forms.
In Section~\ref{sec:examples}, we provide numerical examples, which demonstrate the good conservation properties and long-time stability of the scheme.
In particular, we consider a typical current sheet model as it is used in collisionless reconnection studies and show that reconnection takes place only when electron inertia effects are present but not in the ideal case.

\section{Incompressible Magnetohydrodynamics}\label{sec:ideal_mhd}

The equations of magnetohydrodynamics (MHD) result from the combination of the Navier-Stokes equation for an incompressible fluid with the induction equation of electrodynamics.
In particular, the system of incompressible MHD equations is given by
\begin{subequations}\label{eq:mhd_eqs}
\begin{align}\label{eq:mhd_eqs_V}
\rho \left( V_{t} + ( V \cdot \nabla ) V \right) &= ( \nabla \times B ) \times B + \mu \, \nabla^{2} V - \nabla p , &
\nabla \cdot V &= 0 , \\
\label{eq:mhd_eqs_B}
B_{t} &= \nabla \times ( V \times B ) + \eta \, \nabla^{2} B , &
\nabla \cdot B &= 0 ,
\end{align}
\end{subequations}
where $V$ is the fluid velocity, $B$ is the magnetic field, $p$ is the gas pressure, and $\rho$ is the density, assumed to be constant.
Subscript $t$ denotes the time derivative and the constants $\mu$ and $\eta$ determine the strength of viscosity and resistivity, respectively. The density is set to $\rho = 1$ and the equations are normalised such that the magnetic field $B$ equals the Alfv\'{e}n velocity.

The first equation (\ref{eq:mhd_eqs_V}) is called the \emph{momentum equation}, the second equation (\ref{eq:mhd_eqs_B}) the \emph{induction equation}.
Both $V$ and $B$ are divergence-free, $V$ as we are considering an incompressible fluid, and $B$ as there are no magnetic monopoles.
But while $\nabla \cdot B = 0$ is implied by the induction equation, provided that the initial magnetic field $B(t=0)$ is divergence-free, $\nabla \cdot V = 0$ is a dynamical constraint determining the pressure $p$.

\subsection{Ideal Incompressible Magnetohydrodynamics}

In the following, we will discuss ideal the ideal version of the incompressible MHD equations, that is equations~\eqref{eq:mhd_eqs_V} and~\eqref{eq:mhd_eqs_B} with $\mu = \eta = 0$, thus neglecting viscous as well as resistive effects.
We rewrite the advective derivative of the fluid velocity,
\begin{align}
( V \cdot \nabla ) V = ( \nabla \times V ) \times V + \tfrac{1}{2} \nabla (V \cdot V) ,
\end{align}
so that the the ideal incompressible MHD equations become
\begin{subequations}\label{eq:ideal_mhd_equations}
\begin{align}
\partial_{t} V + \psi(V,V) &= \psi(B,B) - \nabla P , \\
\partial_{t} B + \phi(V,B) &= 0 , \\
\nabla \cdot V &= 0 ,
\end{align}
\end{subequations}
where $P = p + \tfrac{1}{2} \norm{V}^{2}$.
For concise notation, we introduced two bi-linear operators $\psi$ and $\phi$ with components
\begin{subequations}\label{eq:ideal_mhd_equations_operators}
\begin{align}
\psi^{x} (V,B) &\equiv V^{y} \, \big( \partial_{y} B^{x} - \partial_{x} B^{y} \big) , &
\phi^{x} (V,B) &\equiv \partial_{y} \big( V^{y} B^{x} - V^{x} B^{y} \big) , \\
\psi^{y} (V,B) &\equiv V^{x} \, \big( \partial_{x} B^{y} - \partial_{y} B^{x} \big) , &
\phi^{y} (V,B) &\equiv \partial_{x} \big( V^{x} B^{y} - V^{y} B^{x} \big) ,
\end{align}
\end{subequations}
which is the same definitions used by \citeauthor{Gawlik:2011} \cite{Gawlik:2011}.

Neglecting resistivity, $\eta = 0$, equation (\ref{eq:mhd_eqs_B}) states that the magnetic field is advected with the fluid flow, which implies the conservation of the magnetic flux through a surface moving with the fluid \cite{ArnoldKhesin:1998}. In other words, the topology of the magnetic field lines is conserved. They are not allowed to open up and reconnect. A property that is worthwhile to maintain on the discrete level.
In a resistive plasma, $\eta \, \nabla^{2} B$ describes diffusive effects, for which the magnetic field lines are not just dragged along with the field, but are free to change their topology.

As in addition viscosity is neglected, $\mu = 0$, one has three important conserved quantities of ideal MHD in two dimensions \cite{ArnoldKhesin:1998}, namely the total energy, 
\begin{align}\label{eq:ideal_mhd_energy}
E = \dfrac{1}{2} \int \Big[ \norm{V}^{2} + \norm{B}^{2} \Big] \, dx \, dy ,
\end{align}
cross helicity
\begin{align}\label{eq:ideal_mhd_cross_helicity}
C_{\mrm{CH}} = \int V \cdot B \, dx \, dy ,
\end{align}
and magnetic helicity
\begin{align}\label{eq:ideal_mhd_magnetic_helicity}
C_{\mrm{MH}} = \int A \, dx \, dy ,
\end{align}
where $A$ is the vector potential, so that $B = \nabla \times A$.
Conservation of all three quantities is desirable in numerical simulations in order to obtain reliable and physically accurate results.
In the next step, we construct a formal Lagrangian for equations \eqref{eq:ideal_mhd_equations}.

\subsection{Formal Lagrangians}

In order to derive variational integrators for the ideal MHD equations, we need a Lagrangian formulation in Eulerian coordinates.
As such a Lagrangian is not readily available, we have to resort to a formal Lagrangian formulation \cite{KrausMaj:2015, Ibragimov:2006, AthertonHomsy:1975}.
To that end, we treat the ideal MHD system as part of a larger system, which features a Lagrangian formulation.
This approach is described in details in reference \cite{KrausMaj:2015}. Here we outline the procedure for the case at hand without theoretical details.
In practice, each equation of (\ref{eq:ideal_mhd_equations}), including the incompressibility constraint, is multiplied with an auxiliary variable, $\alpha$, $\beta$ and $\gamma$, respectively.
The formal Lagrangian is given as the sum of the resulting expressions,
\begin{align}\label{eq:ideal_mhd_formal_lagrangian}
\mcal{L} (\phy, \phy_{t}, \phy_{x}, \phy_{y})
\nonumber
&= \alpha \cdot \big[ \partial_{t} V + \psi(V,V) - \psi(B,B) + \nabla P \big] \\
&+ \beta  \cdot \big[ \partial_{t} B + \phi(V,B) \big]
 + \gamma \, \big[ \nabla \cdot V \big] .
\end{align}
For concise notation, we write $\phy$ to denote all variables,
\begin{align}
\phy = (V, B, P, \alpha, \beta, \gamma) ,
\end{align}
and $\phy_{t}$, $\phy_{x}$, and $\phy_{y}$ to denote the corresponding derivatives with respect to $t$, $x$ and $y$, respectively.
Requiring stationarity of the action functional (Hamilton's principle), 
\begin{align}
\delta \mcal{A} [\phy] = \delta \int \mcal{L} (\phy, \phy_{t}, \phy_{x}, \phy_{y}) \, dt \, dx \, dy = 0 ,
\end{align}
for variations $\delta \phy$ of the variables, which vanish at the boundaries but are otherwise  arbitrary, gives the ideal MHD equations (\ref{eq:ideal_mhd_equations}) as well as additional equations which determine the evolution of the auxiliary variables,
\begin{subequations}
\begin{align}
\partial_{t} \alpha + \psi (\alpha,V) &= \psi(B, \beta)  + \phi (\alpha,V) - \nabla \gamma , \\
\partial_{t} \beta  + \phi (\alpha,B) &= \psi(\alpha, B) - \psi(V, \beta) , \\
\nabla \cdot \alpha &= 0 .
\end{align}
\end{subequations}
We see that solutions of the equations for the physical variables $V$, $B$ and $P$, will also be solutions of the equations for the auxiliary variables $\alpha$, $\beta$ and $\gamma$, when setting $\alpha(t,x,y) = V(t,x,y)$, $\beta(t,x,y) = B(t,x,y)$ and $\gamma(t,x,y) = P (t,x,y)$.

\subsection{Inertial Magnetohydrodynamics}

We can add effects of electron inertia to the model by introducing a modified magnetic field $\obar{B}$ as follows \cite{Lingam:2015},
\begin{subequations}\label{eq:inertial_mhd_equations}
\begin{align}
\label{eq:inertial_mhd_equations_1}
\partial_{t} V + \psi(V,V) &= \psi(\obar{B}, B) - \nabla P , \\
\label{eq:inertial_mhd_equations_2}
\partial_{t} \obar{B} + \phi(V, \obar{B}) &= 0 , \\
\label{eq:inertial_mhd_equations_3}
\nabla \cdot V &= 0 , \\
\label{eq:inertial_mhd_equations_4}
\obar{B} &= B + d_{e}^{2} ( \nabla \times ( \nabla \times B ) ) ,
\end{align}
\end{subequations}
where $d_{e}$ denotes the electron skin depth. From the last equation, we see that $\nabla \cdot \obar{B} = 0$.
The formal Lagrangian is constructed analogously to \eqref{eq:ideal_mhd_formal_lagrangian}, introducing another auxiliary variable $\sigma$. The only difference is that we integrate the $\sigma \cdot (\nabla \times \nabla \times B)$ term by parts, in order to avoid second order derivatives in the Lagrangian,
\begin{align}\label{eq:inertial_mhd_formal_lagrangian}
\obar{\mcal{L}} (\obar{\phy}, \obar{\phy}_{t}, \obar{\phy}_{x}, \obar{\phy}_{y})
\nonumber
&= \alpha \cdot \big[ \partial_{t} V + \psi(V,V) - \psi(\obar{B},B) + \nabla P \big] \\
\nonumber
&+ \beta  \cdot \big[ \partial_{t} \obar{B} + \phi(V,\obar{B}) \big]
 + \gamma \, \big[ \nabla \cdot V \big] \\
&+ \sigma \cdot \big[ \obar{B} - B \big] - d_{e}^{2} \, \big[ \nabla \times \sigma \big] \cdot \big[ \nabla \times B \big] ,
\end{align}
with $\obar{\phy}$ the extended solution vector
\begin{align}
\obar{\phy} = (V, \obar{B}, P, B, \alpha, \beta, \gamma, \sigma) .
\end{align}
The inertial MHD system has a modified set of conservation laws, namely energy
\begin{align}\label{eq:inertial_mhd_energy}
\obar{E} = \dfrac{1}{2} \int \Big[ \norm{V}^{2} + B \cdot \obar{B} \Big] \, dx \, dy ,
\end{align}
cross helicity
\begin{align}\label{eq:inertial_mhd_cross_helicity}
\obar{C}_{\mrm{CH}} = \int V \cdot \obar{B} \, dx \, dy ,
\end{align}
and magnetic helicity
\begin{align}\label{eq:inertial_mhd_magnetic_helicity}
\obar{C}_{\mrm{MH}} = \int \obar{A} \, dx \, dy ,
\end{align}
where $\obar{A}$ is the generalised vector potential, so that $\obar{B} = \nabla \times \obar{A}$.

\section{Variational Discretisation}\label{sec:vi}

In order to obtain a numerical method for the ideal MHD equations (\ref{eq:ideal_mhd_equations}), we discretise the action functional $\mcal{A}$ and apply a discrete version of Hamilton's principle of stationary action \cite{MarsdenPatrick:1998, KrausMaj:2015, KrausMaj:2017}.
Standard variational discretisations on cartesian meshes like Veselov-type discretisations or the box scheme usually lead to centred finite difference schemes, which are problematic for Euler's equation.
Such schemes are known to be prone to instabilities referred to as \emph{checker-boarding} (see e.g.~\citet{Langtangen:2002} or~\citet{McDonough:2007}), originating from co-locating the components of the velocity vector and the pressure at the same grid points.
This often leads to solutions with highly oscillatory pressure as symmetric difference operators for the gradient, e.g., with stencil $[ -1 \; \hphantom{-}0 \; +1 ]$, annihilate pressures which oscillate between neighbouring grid points, e.g., between~$-1$ and~$+1$.
With finite difference methods, this is usually circumvented by introducing a staggered grid, where only the pressure at a single grid point enforces the divergence of the velocity of the surrounding grid points to vanish (c.f. Figure~\ref{fig:mhd_staggered_grid_div}).

The discretisation described next follows such an approach and is based on discrete differential forms as is explained in detail by~\citet{KrausMaj:2017}. Here, we employ a more heuristic derivation that is accessible also without an in-depth background in differential forms and discrete differential geometry. It is worthwhile to note, though, that all of what is presented in the next section follows from a rigorous framework.

\subsection{Staggered Grid}\label{sec:staggered_grid}

\begin{figure}[tb]
\centering
\subfloat[Momentum and Induction Equation]{\label{fig:mhd_staggered_grid_momentum}
\includegraphics[width=.44\textwidth]{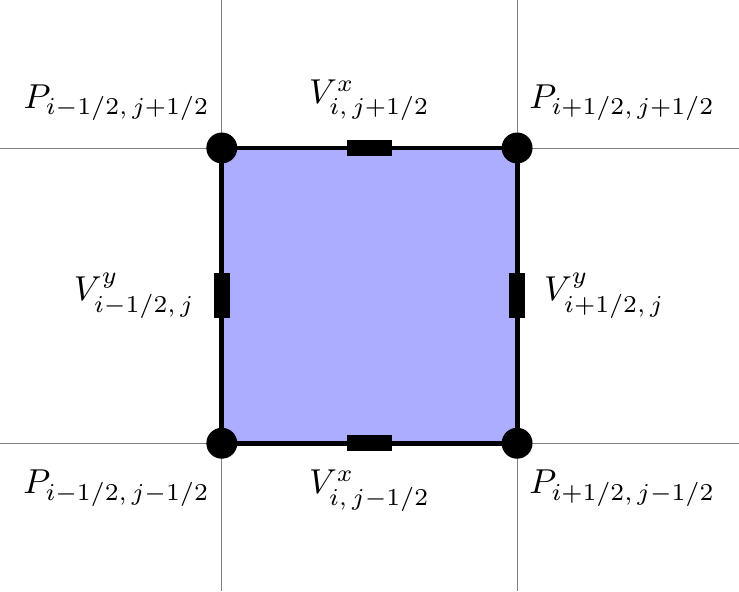}
}
\subfloat[Divergence Constraint]{\label{fig:mhd_staggered_grid_div}
\includegraphics[width=.44\textwidth]{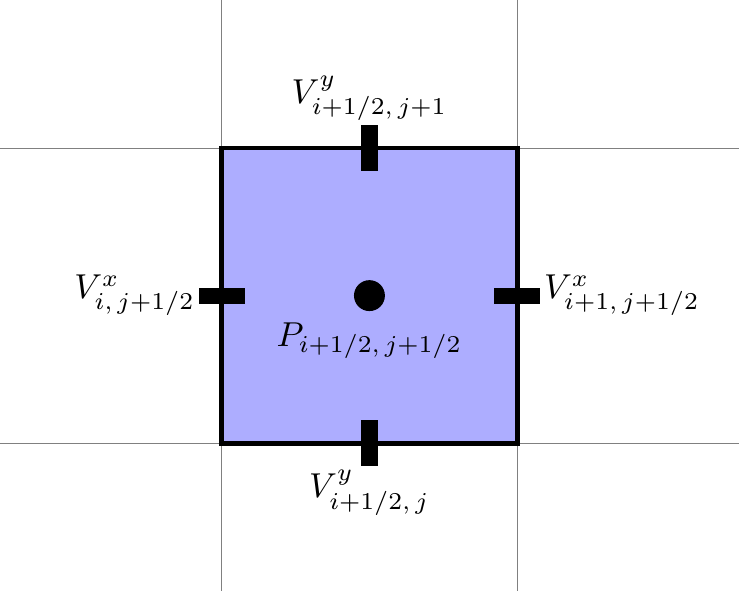}
}
\caption{Staggered grid in the $xy$-plane. Left: Primal grid with natural positions for the pressure and the velocity components for the computation of the advection operators. Right: Dual grid for the computation of the divergence constraint.}
\label{fig:mhd_staggered_grid}
\end{figure}

We introduce a staggered grid, where the pressure is collocated at the vertices of a grid cell and the velocity components at the edges, like it is depicted in Figure~\subref*{fig:mhd_staggered_grid_momentum}.
The location of the physical quantities comes natural when viewed as differential forms. The pressure is a zero-form and is therefore collocated at the vertices of a cell of the primal grid. The velocity (and in two dimensions also the magnetic field) is a one-form and is therefore collocated at the edges of a cell, $x$-components on the horizontal edges and $y$-components on the vertical edges (c.f. Figure~\subref*{fig:mhd_staggered_grid_momentum}).

On the dual grid, the pressure becomes a two-form, collocated at the cell centre. The velocity and magnetic field are still one-forms, but twisted, so that $x$-components are collocated on the vertical edges and $y$-components on the horizontal edges (c.f. Figure~\subref*{fig:mhd_staggered_grid_div}).
This can also be seen by considering the discrete divergence-free constraint of the velocity field,
\begin{align}\label{eq:mhd_staggered_grid_divergence}
  \dfrac{V^{x}_{i, \, j+1/2, \, n} - V^{x}_{i-1, \, j+1/2, \, n}}{h_{x}}
+ \dfrac{V^{y}_{i+1/2, \, j, \, n} - V^{y}_{i+1/2, \, j-1, \, n}}{h_{y}} = 0 ,
\end{align}
which is defined in such a way that the natural location of the divergence coincides with the location of the pressure.
This is important as the function of the pressure in incompressible fluid dynamics can be described as enforcing the divergence-free constraint of the velocity field.

On the primal grid, we define point-wise discrete time-derivatives,
\begin{subequations}
\begin{alignat}{5}
& \partial_{t} V^{x} &
& \quad \rightarrow \quad &
& ( \Delta_{t} V^{x} )_{i, \, j+1/2, \, n+1/2} &
& \equiv \; &
& \dfrac{V^{x}_{i, \, j+1/2, \, n+1} - V^{x}_{i, \, j+1/2, \, n}}{h_{t}} ,
\\
& \partial_{t} V^{y} &
& \quad \rightarrow \quad &
& ( \Delta_{t} V )^{y}_{i+1/2, \, j, \, n+1/2} &
& \equiv \; &
& \dfrac{V^{y}_{i+1/2, \, j, \, n+1} - V^{y}_{i+1/2, \, j, \, n}}{h_{t}} .
\end{alignat}
\end{subequations}
For the spatial derivatives of the vectors, we use midpoint averaging with respect to time.
The $x$-derivative of $x$-components and the $y$-derivative of $y$-components are defined on the dual grid as
\begingroup
\allowdisplaybreaks
\begin{subequations}\label{eq:discrete_field_derivatives_1}
\begin{alignat}{9}
\nonumber
& \partial_{x} V^{x} &
& \quad \rightarrow \quad &
& ( \Delta_{x} V^{x} )_{i+1/2, \, j+1/2, \, n+1/2} &
& \equiv \; &
& \dfrac{1}{2} \, \bigg[ \dfrac{V^{x}_{i+1, \, j+1/2, \, n  } - V^{x}_{i, \, j+1/2, \, n  }}{h_{x}}
\\
&&&&&&&&& \hspace{2em}
+ \dfrac{V^{x}_{i+1, \, j+1/2, \, n+1} - V^{x}_{i, \, j+1/2, \, n+1}}{h_{x}} \bigg] ,
\\
\nonumber
& \partial_{y} V^{y} &
& \quad \rightarrow \quad &
& ( \Delta_{y} V^{y} )_{i+1/2, \, j+1/2, \, n+1/2} &
& \equiv \; &
& \dfrac{1}{2} \, \bigg[ \dfrac{V^{y}_{i+1/2, \, j+1, \, n  } - V^{y}_{i+1/2, \, j, \, n  }}{h_{y}}
\\
&&&&&&&&& \hspace{2em}
+ \dfrac{V^{y}_{i+1/2, \, j+1, \, n+1} - V^{y}_{i+1/2, \, j, \, n+1}}{h_{y}} \bigg] ,
\end{alignat}
while the $y$-derivative of $x$-components and the $x$-derivative of $y$-components are defined on the primal grid as
\begin{alignat}{9}
\nonumber
& \partial_{y} V^{x} &
& \quad \rightarrow \quad &
& ( \Delta_{y} V^{x} )_{i, \, j, \, n+1/2} &
& \equiv \; &
& \dfrac{1}{2} \, \bigg[ \dfrac{V^{x}_{i, \, j+1/2, \, n  } - V^{x}_{i, \, j-1/2, \, n  }}{h_{y}}
\\
&&&&&&&&& \hspace{2em}
+ \dfrac{V^{x}_{i, \, j+1/2, \, n+1} - V^{x}_{i, \, j-1/2, \, n+1}}{h_{y}} \bigg] ,
\end{alignat}
\end{subequations}
\begin{subequations}\label{eq:discrete_field_derivatives_2}
\begin{alignat}{9}
\nonumber
& \partial_{x} V^{y} &
& \quad \rightarrow \quad &
& ( \Delta_{x} V^{y} )_{i, \, j, \, n+1/2} &
& \equiv \; &
& \dfrac{1}{2} \, \bigg[ \dfrac{V^{y}_{i+1/2, \, j, \, n  } - V^{y}_{i-1/2, \, j, \, n  }}{h_{x}} \\
&&&&&&&&& \hspace{2em}
+ \dfrac{V^{y}_{i+1/2, \, j, \, n+1} - V^{y}_{i-1/2, \, j, \, n+1}}{h_{x}} \bigg] ,
\end{alignat}
\end{subequations}
\endgroup
Note, that the $x$-derivative of $V^{x}$ and the $y$-derivative of $V^{y}$ are defined on the grid in Figure~\subref*{fig:mhd_staggered_grid_div}, while the $y$-derivative of $V^{x}$ and the $x$-derivative of $V^{y}$ are defined on the dual grid in Figure~\subref*{fig:mhd_staggered_grid_momentum}.
The indices of the derivatives denote the natural collocation of the derivative, which is always the cell centre.

Derivatives of the pressure can only be defined on the primal grid (c.f.~Figure~\subref*{fig:mhd_staggered_grid_momentum}).
They are naturally defined along the edges of the cells.
The staggering approach is applied to $P$ also with respect to time, i.e., the pressure nodes are $(i+1/2, \, j+1/2, \, n+1/2)$.
Taking all of this into account, we define
\begin{subequations}
\begin{alignat}{9}
& \partial_{x} P && \quad \rightarrow \quad &
& ( \Delta_{x} P )_{i, \, j+1/2, \, n+1/2} &
& \equiv \; &
& \dfrac{P_{i+1/2, \, j+1/2, \, n+1/2} - P_{i-1/2, \, j+1/2, \, n+1/2}}{h_{x}} ,
\\
& \partial_{y} P && \quad \rightarrow \quad &
& ( \Delta_{y} P )_{i+1/2, \, j, \, n+1/2} &
& \equiv \; &
& \dfrac{P_{i+1/2, \, j+1/2, \, n+1/2} - P_{i+1/2, \, j-1/2, \, n+1/2}}{h_{y}} .
\end{alignat}
\end{subequations}
Averages of the vector fields are only needed on the primal grid, so we are defining them only there. For $V$ and $B$, the averaging is applied with respect to both space and time,
\begin{subequations}
\begin{align}
\bracketb{ V^{x} }_{i, \, j, \, n+1/2}
&\equiv \dfrac{1}{4} \Big[
  V^{x}_{ i, \, j-1/2 , \, n   } + V^{x}_{ i, \, j+1/2, \, n   }
+ V^{x}_{ i, \, j-1/2 , \, n+1 } + V^{x}_{ i, \, j+1/2, \, n+1 } \Big] ,
\\
\bracketb{ V^{y} }_{i, \, j, \, n+1/2}
&\equiv \dfrac{1}{4} \Big[
  V^{y}_{ i-1/2 , \, j, \, n   } + V^{y}_{ i+1/2, \, j, \, n   }
+ V^{y}_{ i-1/2 , \, j, \, n+1 } + V^{y}_{ i+1/2, \, j, \, n+1 } \Big] ,
\end{align}
\end{subequations}
but as $\alpha$ and $\beta$ will be collocated at $n+1/2$ (see comment in the next section), their averages do not involve time, but only space, in particular
\begin{subequations}
\begin{align}
\bracketb{ \alpha^{x} }_{i, \, j, \, n+1/2}
&\equiv \dfrac{1}{2} \Big[
 \alpha^{x}_{ i, \, j-1/2 , \, n+1/2 } + \alpha^{x}_{ i, \, j+1/2, \, n+1/2 } \Big] ,
\\
\bracketb{ \alpha^{y} }_{i, \, j, \, n+1/2}
&\equiv \dfrac{1}{2} \Big[
 \alpha^{y}_{ i-1/2 , \, j, \, n+1/2 } + \alpha^{y}_{ i+1/2, \, j, \, n+1/2 } \Big] .
\end{align}
\end{subequations}
With these definitions we will now construct the discrete Lagrangians.

\subsection{Euler Equation}

We start the derivation of the variational integrator by considering the incompressible Euler equation,
\begin{align}
\partial_{t} V + \psi (V,V) - \psi(B,B) + \nabla P &= 0 , &
\nabla \cdot V &= 0 .
\end{align}
The action integral of the formal Lagrangian (\ref{eq:ideal_mhd_formal_lagrangian}), reduced to this subsystem, is
\begin{align}\label{eq:mhd_navier_stokes_lagrangian}
\mcal{A} &= \int \Big[ ... + \alpha \cdot \big[ \partial_{t} V + \psi(V,V) - \psi(B,B) + \nabla P \big] + \gamma \, \big[ \nabla \cdot V \big] + ... \Big] \, dt \, dx \, dy .
\end{align}
To be able to discretise all of the derivatives in the first term of the Lagrangian, we have to use the primal grid, as depicted in Figure~\subref*{fig:mhd_staggered_grid_momentum}.
The time derivatives are approximated using the trapezoidal rule in space,
\begin{subequations}
\begin{align}
\alpha^{x} \, \partial_{t} V^{x} \quad \rightarrow \quad
\dfrac{1}{2} \Big[
\nonumber
& \alpha^{x}_{ i, \, j-1/2, \, n+1/2 } \, ( \Delta_{t} V^{x} )_{i, \, j-1/2, \, n+1/2} \\
& \qquad
+ \alpha^{x}_{ i, \, j+1/2, \, n+1/2 } \, ( \Delta_{t} V^{x} )_{i, \, j+1/2, \, n+1/2} 
  \Big] , \\
\alpha^{y} \, \partial_{t} V^{y} \quad \rightarrow \quad
\dfrac{1}{2} \Big[
\nonumber
& \alpha^{y}_{ i-1/2, \, j, \, n+1/2 } \, ( \Delta_{t} V^{y} )_{i-1/2, \, j, \, n+1/2} \\
& \qquad
+ \alpha^{y}_{ i+1/2, \, j, \, n+1/2 } \, ( \Delta_{t} V^{y} )_{i+1/2, \, j, \, n+1/2} 
  \Big] .
\end{align}
\end{subequations}
The multiplier $\alpha$ is collocated at $n+1/2$, \, just as the time derivative.
We use a trapezoidal approximation to avoid spatial averaging of the time derivatives in the resulting scheme, as that might lead to grid oscillations (checker-boarding) in the velocity field.
We apply the same approximation to the pressure gradient term, for the same reason, namely
\begin{subequations}
\begin{align}
\alpha^{x} \, \partial_{x} P \quad \rightarrow \quad
\dfrac{1}{2} \Big[
\nonumber
& \alpha^{x}_{ i, \, j-1/2, \, n+1/2 } \, ( \Delta_{x} P )_{i, \, j-1/2, \, n+1/2} \\
& \qquad
+ \alpha^{x}_{ i, \, j+1/2, \, n+1/2 } \, ( \Delta_{x} P )_{i, \, j+1/2, \, n+1/2} 
  \Big] ,
\\
\alpha^{y} \, \partial_{y} P \quad \rightarrow \quad
\dfrac{1}{2} \Big[
\nonumber
& \alpha^{y}_{ i-1/2, \, j, \, n+1/2 } \, ( \Delta_{x} P )_{i-1/2, \, j, \, n+1/2} \\
& \qquad
+ \alpha^{y}_{ i+1/2, \, j, \, n+1/2 } \, ( \Delta_{y} P )_{i+1/2, \, j, \, n+1/2} 
  \Big] .
\end{align}
\end{subequations}
As previously mentioned, the pressure is collocated at $n+1/2$, such that no time average of $P$ is needed.
The $\psi$ operator~(\ref{eq:ideal_mhd_equations_operators}) is discretised by a midpoint approximation, both with respect to space and time, i.e.,
\begin{subequations}\label{eq:mhd_discrete_psi_action}
\begin{align}
\psi^{x} (V,V) \quad \rightarrow \quad
\psi^{x}_{i, \, j, \, n+1/2} (V,V)
\equiv \bracketb{ V^{y} }_{i, \, j, \, n+1/2} \, \Big[
\nonumber
& ( \Delta_{y} V^{x} )_{i, \, j, \, n+1/2} \\
& \qquad
- ( \Delta_{x} V^{y} )_{i, \, j, \, n+1/2}
\Big] ,
\\
\psi^{y} (V,V) \quad \rightarrow \quad
\psi^{y}_{i, \, j, \, n+1/2} (V,V)
\equiv \bracketb{ V^{x} }_{i, \, j, \, n+1/2} \, \Big[
\nonumber
& ( \Delta_{x} V^{y} )_{i, \, j, \, n+1/2} \\
& \qquad
- ( \Delta_{y} V^{x} )_{i, \, j, \, n+1/2}
\Big] ,
\end{align}
\end{subequations}
so that
\begin{subequations}
\begin{align}
\alpha^{x} \psi^{x} (V,V) \quad \rightarrow \quad
& \bracketb{ \alpha^{x} }_{i, \, j, \, n+1/2} \, \psi^{x}_{i, \, j, \, n+1/2} (V,V) ,
\\
\alpha^{y} \psi^{y} (V,V) \quad \rightarrow \quad
& \bracketb{ \alpha^{y} }_{i, \, j, \, n+1/2} \, \psi^{y}_{i, \, j, \, n+1/2} (V,V) ,
\end{align}
\end{subequations}
and analogously for the magnetic force term $\psi(B,B)$.

The discretisation of the divergence term in~(\ref{eq:mhd_navier_stokes_lagrangian}) is implemented on the dual grid in Figure~\subref*{fig:mhd_staggered_grid_div}.
Recognising that $\gamma$ is a scalar field and thus collocated at the same position as the pressure, the discretisation follows directly from (\ref{eq:mhd_staggered_grid_divergence}), i.e.,
\begin{align}
\gamma \, \big( \nabla \cdot V \big) \quad \rightarrow \quad
\gamma_{ i+1/2, \, j+1/2, \, n } \, \Big[
  ( \Delta_{x} V^{x} )_{i+1/2, \, j+1/2, \, n}
+ ( \Delta_{y} V^{y} )_{i+1/2, \, j+1/2, \, n}
\Big] .
\end{align}
Summing up all contributions yields the discrete Lagrangian for the momentum equation,
\begin{align}
L^{\text{M}}_{i,j,n+1/2}
\nonumber
= h_{t} h_{x} h_{y} \bigg\{
& \dfrac{1}{2} \Big[
     \alpha^{x}_{ i, \, j-1/2, \, n+1/2 } \, ( \Delta_{t} V^{x} )_{i, \, j-1/2, \, n+1/2}
   + \alpha^{x}_{ i, \, j+1/2, \, n+1/2 } \, ( \Delta_{t} V^{x} )_{i, \, j+1/2, \, n+1/2} 
   \Big] \\
\nonumber
+&{} \dfrac{1}{2} \Big[
      \alpha^{y}_{ i-1/2, \, j, \, n+1/2 } \, ( \Delta_{t} V^{y} )_{i-1/2, \, j, \, n+1/2}
    + \alpha^{y}_{ i+1/2, \, j, \, n+1/2 } \, ( \Delta_{t} V^{y} )_{i+1/2, \, j, \, n+1/2} 
   \Big] \\
\nonumber
+&{} \dfrac{1}{2} \Big[
      \alpha^{x}_{ i, \, j-1/2, \, n+1/2 } \, ( \Delta_{x} P )_{i, \, j-1/2, \, n+1/2}
    + \alpha^{x}_{ i, \, j+1/2, \, n+1/2 } \, ( \Delta_{x} P )_{i, \, j+1/2, \, n+1/2} 
   \Big] \\
\nonumber
+&{} \dfrac{1}{2} \Big[
      \alpha^{y}_{ i-1/2, \, j, \, n+1/2 } \, ( \Delta_{x} P )_{i-1/2, \, j, \, n+1/2}
    + \alpha^{y}_{ i+1/2, \, j, \, n+1/2 } \, ( \Delta_{y} P )_{i+1/2, \, j, \, n+1/2} 
   \Big] \\
\nonumber
\vphantom{\dfrac{1}{2}}
+&{} \bracketb{ \alpha^{x} }_{i, \, j, \, n+1/2} \, \psi^{x}_{i, \, j, \, n+1/2} (V,V)
-    \bracketb{ \alpha^{x} }_{i, \, j, \, n+1/2} \, \psi^{x}_{i, \, j, \, n+1/2} (B,B) \\
\vphantom{\dfrac{1}{2}}
+&{} \bracketb{ \alpha^{y} }_{i, \, j, \, n+1/2} \, \psi^{y}_{i, \, j, \, n+1/2} (V,V)
-    \bracketb{ \alpha^{y} }_{i, \, j, \, n+1/2} \, \psi^{y}_{i, \, j, \, n+1/2} (B,B)
\bigg\} .
\end{align}
In addition we obtain the Lagrangian for the divergence constraint,
\begin{align}
L^{\text{D}}_{i,j,n}
&= h_{t} h_{x} h_{y} \, \gamma_{ i+1/2, \, j+1/2, \, n } \, \Big[
       ( \Delta_{x} V^{x} )_{i+1/2, \, j+1/2, \, n}
     + ( \Delta_{y} V^{y} )_{i+1/2, \, j+1/2, \, n}
\Big] ,
\end{align}
which in contrast to $L^{\text{M}}_{i,j,n+1/2}$ is not defined at the midpoint but at integer times.
We can, however, define
\begin{align}
L^{\text{D}}_{i,j,n+1/2} = \dfrac{1}{2} \Big[ L^{\text{D}}_{i,j,n} + L^{\text{D}}_{i,j,n+1} \Big] ,
\end{align}
so that all Lagrangians are collocated at the same point in space as well as in time.

\subsection{Induction Equation}

Now we consider those terms of the action that will yield the induction equation, i.e.,
\begin{align}
\mcal{A} &= \int \Big[ ... + \beta \cdot \big[ \partial_{t} B + \phi(V,B) \big] + ... \Big] \, dt \, dx \, dy .
\end{align}
To find a discretisation of the $\phi(V,B)$ operator~(\ref{eq:ideal_mhd_equations_operators}) on a single grid cell, we have to perform an integration by parts, such that
\begin{align}
\mcal{A}
\nonumber
&= \int \Big[ ... + \beta^{x} \, \big[ \partial_{t} B^{x} - \partial_{y} ( V^{x} B^{y} - V^{y} B^{x} ) \big] \\
\nonumber
& \hspace{8em}
+ \beta^{y} \, \big[ \partial_{t} B^{y} + \partial_{x} ( V^{x} B^{y} - V^{y} B^{x} ) \big] + ... \Big] \, dt \, dx \, dy \\
\nonumber
&= \int \Big[ ... + \beta^{x} \, \partial_{t} B^{x} + ( \partial_{y} \beta^{x} ) ( V^{x} B^{y} - V^{y} B^{x} ) \\
& \hspace{8em}
+ \beta^{y} \, \partial_{t} B^{y} - ( \partial_{x} \beta^{y} ) ( V^{x} B^{y} - V^{y} B^{x} ) + ... \Big] \, dt \, dx \, dy
,
\end{align}
assuming appropriate boundary conditions, such that the boundary terms vanish (e.g.,~periodic or homogeneous Dirichlet).
The discretisation of the time derivative is the same as in the case of the momentum equation, i.e., using the trapezoidal rule in space,
\begin{subequations}
\begin{align}
\beta^{x} \, \partial_{t} B^{x} \quad \rightarrow \quad
\dfrac{1}{2} \Big[
\nonumber
& \beta^{x}_{ i, \, j-1/2, \, n+1/2 } \, ( \Delta_{t} B^{x} )_{i, \, j-1/2, \, n+1/2} \\
& \qquad
+ \beta^{x}_{ i, \, j+1/2, \, n+1/2 } \, ( \Delta_{t} B^{x} )_{i, \, j+1/2, \, n+1/2} 
  \Big] , \\
\beta^{y} \, \partial_{t} B^{y} \quad \rightarrow \quad
 \dfrac{1}{2} \Big[
\nonumber
& \beta^{y}_{ i-1/2, \, j, \, n+1/2 } \, ( \Delta_{t} B^{y} )_{i-1/2, \, j, \, n+1/2} \\
& \qquad
+ \beta^{y}_{ i+1/2, \, j, \, n+1/2 } \, ( \Delta_{t} B^{y} )_{i+1/2, \, j, \, n+1/2} 
  \Big] .
\end{align}
\end{subequations}
The factors of the operator  $\phi$ are collocated at different positions of the grid, i.e., $\Delta_{y} \beta^{x}$ and $\Delta_{x} \beta^{y}$ are collocated at $(i, \, j)$, while $V^{x}$ and $B^{x}$ are collocated at $(i, \, j+1/2)$, and $V^{y}$ and $B^{y}$ are collocated at $(i+1/2, \, j)$.
In order to compute products of these expressions, all factors should be collocated at the same position, namely the cell centres $(i,j)$.
Therefore, we multiply $\Delta_{y} \beta^{x}$ and $\Delta_{x} \beta^{y}$, which are already collocated at the cell centres, with midpoint averages of the vector fields $V$ and $B$, that is
\begin{subequations}
\begin{align}
\nonumber
( \partial_{y} \beta^{x} ) ( V^{x} B^{y} - V^{y} B^{x} ) \quad \rightarrow \quad
  \Delta_{y} \beta^{x}_{i, \, j, \, n+1/2}
\, \Big[
& \bracketb{ V^{x} }_{i, \, j, \, n+1/2} \, \bracketb{ B^{y} }_{i, \, j, \, n+1/2} \\
& \qquad
- \bracketb{ V^{y} }_{i, \, j, \, n+1/2} \, \bracketb{ B^{x} }_{i, \, j, \, n+1/2}
\Big] ,
\\
\nonumber
( \partial_{x} \beta^{y} ) ( V^{x} B^{y} - V^{y} B^{x} ) \quad \rightarrow \quad
  \Delta_{x} \beta^{y}_{i, \, j, \, n+1/2}
\, \Big[
& \bracketb{ V^{x} }_{i, \, j, \, n+1/2} \, \bracketb{ B^{y} }_{i, \, j, \, n+1/2} \\
& \qquad
- \bracketb{ V^{y} }_{i, \, j, \, n+1/2} \, \bracketb{ B^{x} }_{i, \, j, \, n+1/2}
\Big] .
\end{align}
\end{subequations}
Putting all terms together, we obtain the discrete Lagrangian for the induction equation,
\begin{align}
L^{\text{I}}_{i,j,n+1/2}
= h_{t} h_{x} h_{y} \bigg\{ 
\nonumber
& \dfrac{1}{2} \Big[
  \beta^{x}_{ i, \, j-1/2, \, n+1/2 } \, ( \Delta_{t} B^{x} )_{i, \, j-1/2, \, n+1/2} \\
\nonumber
& \hspace{8em}
+ \beta^{x}_{ i, \, j+1/2, \, n+1/2 } \, ( \Delta_{t} B^{x} )_{i, \, j+1/2, \, n+1/2} 
  \Big] \\
\nonumber
+&{} \dfrac{1}{2} \Big[
  \beta^{y}_{ i-1/2, \, j, \, n+1/2 } \, ( \Delta_{t} B^{y} )_{i-1/2, \, j, \, n+1/2} \\
\nonumber
& \hspace{8em}
+ \beta^{y}_{ i+1/2, \, j, \, n+1/2 } \, ( \Delta_{t} B^{y} )_{i+1/2, \, j, \, n+1/2} 
  \Big] \\
\nonumber
-&{} ( \Delta_{y} \beta^{x} )_{i, \, j, \, n+1/2} \, \Big[
     \bracketb{ V^{y} }_{i, \, j, \, n+1/2} \, \bracketb{ B^{x} }_{i, \, j, \, n+1/2} \\
\nonumber
& \hspace{8em}
   - \bracketb{ V^{x} }_{i, \, j, \, n+1/2} \, \bracketb{ B^{y} }_{i, \, j, \, n+1/2}
\Big] \\
\nonumber
-&{} ( \Delta_{x} \beta^{y} )_{i, \, j, \, n+1/2} \, \Big[
     \bracketb{ V^{x} }_{i, \, j, \, n+1/2} \, \bracketb{ B^{y} }_{i, \, j, \, n+1/2} \\
& \hspace{8em}
   - \bracketb{ V^{y} }_{i, \, j, \, n+1/2} \, \bracketb{ B^{x} }_{i, \, j, \, n+1/2}
\Big]
\bigg\} .
\end{align}
Now we have all the ingredients for a complete discretisation of the action integral corresponding to~(\ref{eq:ideal_mhd_equations}).

\subsection{Variational Integrator}

\begin{figure}[bt]
\centering
\includegraphics[width=.44\textwidth]{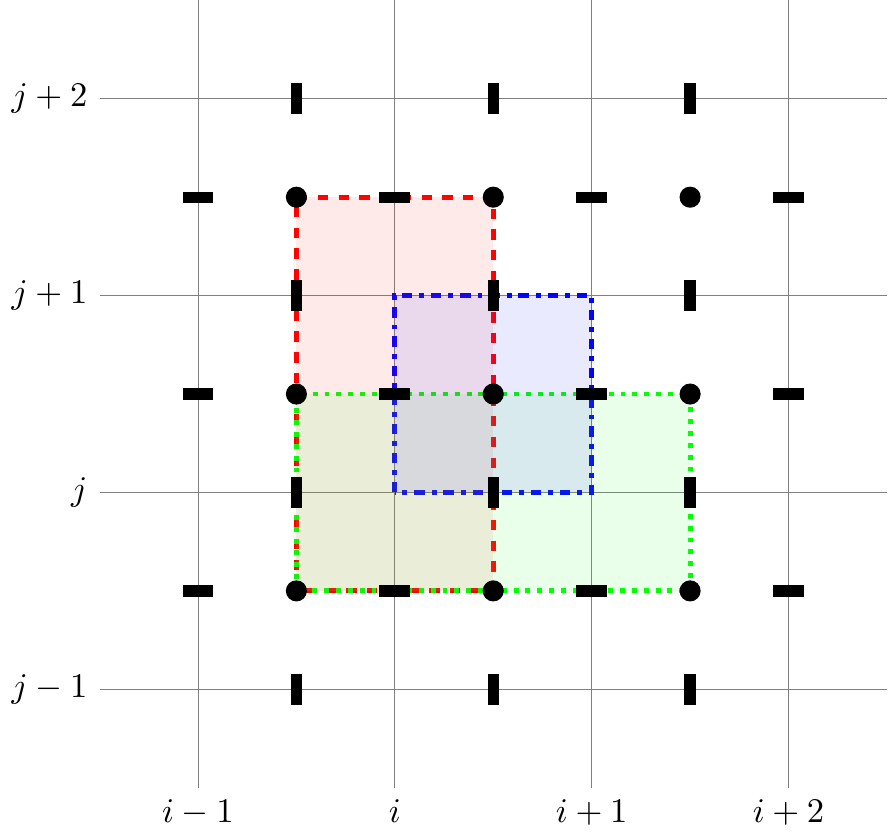}
\caption{Stencils on the staggered grid in the $xy$-plane: $x$-component (red) and $y$-component (green) of momentum and induction equation, divergence constraint (blue).}
\label{fig:mhd_staggered_grid_scheme}
\end{figure}

Requiring stationarity of the discrete action, 
\begin{align}
\delta \mcal{A}_{d} [\phy_{d}] = \delta \sum \limits_{i=1}^{n_{x}} \sum \limits_{j=1}^{n_{y}} \sum \limits_{n=0}^{n_{t}-1} \Big[ L^{\text{M}}_{i,j,n+1/2} + L^{\text{D}}_{i,j,n+1/2} + L^{\text{I}}_{i,j,n+1/2} \Big] = 0 , 
\end{align}
yields the discrete ideal MHD equations,
\begingroup
\allowdisplaybreaks
\begin{subequations}\label{eq:variational_integrator_ideal_mhd}
\begin{align}
\nonumber
& ( \Delta_{t} V^{x} )_{i, \, j+1/2, \, n+1/2}
+ \psi^{x}_{i, \, j+1/2, \, n+1/2} (V,V) \\
& \hspace{8em}
- \psi^{x}_{i, \, j+1/2, \, n+1/2} (B,B)
+ ( \Delta_{x} P )_{i, \, j+1/2, \, n+1/2} = 0 ,
\\
\nonumber
& ( \Delta_{t} V^{y} )_{i+1/2, \, j, \, n+1/2}
+ \psi^{y}_{i+1/2, \, j, \, n+1/2} (V,V) \\
& \hspace{8em}
- \psi^{y}_{i+1/2, \, j, \, n+1/2} (B,B)
+ ( \Delta_{y} P_{i+1/2, \, j, \, n+1/2} = 0 ,
\\
& ( \Delta_{t} B^{x} )_{i, \, j+1/2, \, n+1/2}
+ \phi^{x}_{i, \, j+1/2, \, n+1/2} (V,B) = 0 ,
\\
& ( \Delta_{t} B^{y} )_{i+1/2, \, j, \, n+1/2}
+ \phi^{y}_{i+1/2, \, j, \, n+1/2} (V,B) = 0 ,
\\
&  ( \Delta_{x} V^{x} )_{i+1/2, \, j+1/2, \, n+1/2}
 + ( \Delta_{y} V^{y} )_{i+1/2, \, j+1/2, \, n+1/2} = 0 .
\end{align}
\end{subequations}
\endgroup
Here, $\phy_{d}$ denotes the discrete solution,
\begin{align}
\phy_{d} = \Big\{
\nonumber
& V^{x}_{i,j+1/2,n}, \, V^{y}_{i+1/2,j,n}, \, B^{x}_{i,j+1/2,n}, \, B^{y}_{i+1/2,j,n}, \, P_{i+1/2,j+1/2,m+1/2}, \\
\nonumber
& \alpha^{x}_{i,j+1/2,m+1/2}, \, \alpha^{y}_{i+1/2,j,m+1/2}, \, \beta^{x}_{i,j+1/2,m+1/2}, \, \beta^{y}_{i+1/2,j,m+1/2}, \, \gamma_{i+1/2,j+1/2,n} \\
& \Big\vert \; 1 \leq i \leq n_{x} , \, 1 \leq j \leq n_{y} , \, 0 \leq m < n_{t} - 1 , \, 0 \leq n \leq n_{t} \Big\} ,
\end{align}
where in the following we are considering a periodic domain, so that e.g. the index $n_{x} + 1/2$ denotes the same grid point as the index $1/2$, etc.,
and the discrete operators are defined by
\begin{subequations}
\begin{align}
\psi^{x}_{i, \, j+1/2, \, n+1/2} (V,V)
\nonumber
&= \dfrac{1}{2} \Big[ \psi^{x}_{i, \, j, \, n+1/2} (V,V) + \psi^{x}_{i, \, j+1, \, n+1/2} (V,V) \Big] \\
\nonumber
&= \dfrac{1}{2} \bracketb{ V^{x} }_{i, \, j,   \, n+1/2} \Big[
     \Delta_{y} V^{x}_{i, \, j,   \, n+1/2}
   - \Delta_{x} V^{y}_{i, \, j,   \, n+1/2}
\Big] \\
& \qquad
 + \dfrac{1}{2} \bracketb{ V^{x} }_{i, \, j+1, \, n+1/2} \Big[
     \Delta_{y} V^{x}_{i, \, j+1, \, n+1/2}
   - \Delta_{x} V^{y}_{i, \, j+1, \, n+1/2}
\Big] , \\
\psi^{y}_{i+1/2, \, j, \, n+1/2} (V,V)
\nonumber
&= \dfrac{1}{2} \Big[ \psi^{y}_{i, \, j, \, n+1/2} (V,V) + \psi^{y}_{i+1, \, j, \, n+1/2} (V,V) \Big] \\
\nonumber
&= \dfrac{1}{2} \bracketb{ V^{y} }_{i,   \, j, \, n+1/2} \Big[
     \Delta_{x} V^{y}_{i,   \, j, \, n+1/2}
   - \Delta_{y} V^{x}_{i,   \, j, \, n+1/2}
\Big] \\
& \qquad
 + \dfrac{1}{2} \bracketb{ V^{y} }_{i+1, \, j, \, n+1/2} \Big[
     \Delta_{x} V^{y}_{i+1, \, j, \, n+1/2}
   - \Delta_{y} V^{x}_{i+1, \, j, \, n+1/2}
\Big] ,
\end{align}
\end{subequations}
and
\begin{subequations}
\begin{align}
\phi^{x}_{i, \, j+1/2, \, n+1/2} (V,B)
\nonumber
&= \dfrac{1}{2} \Big[
       \bracketb{ V^{x} }_{i, \, j+1, \, n+1/2} \, \bracketb{ B^{y} }_{i, \, j+1, \, n+1/2}
     - \bracketb{ V^{y} }_{i, \, j+1, \, n+1/2} \, \bracketb{ B^{x} }_{i, \, j+1, \, n+1/2}
\\
& \hspace{1.5em}
     - \bracketb{ V^{x} }_{i, \, j, \, n+1/2} \, \bracketb{ B^{y} }_{i, \, j, \, n+1/2}
     - \bracketb{ V^{y} }_{i, \, j, \, n+1/2} \, \bracketb{ B^{x} }_{i, \, j, \, n+1/2}
\Big] ,
\\
\phi^{y}_{i+1/2, \, j, \, n+1/2} (V,B)
\nonumber
&= \dfrac{1}{2} \Big[
       \bracketb{ V^{y} }_{i+1, \, j, \, n+1/2} \, \bracketb{ B^{x} }_{i+1, \, j, \, n+1/2}
     - \bracketb{ V^{x} }_{i+1, \, j, \, n+1/2} \, \bracketb{ B^{y} }_{i+1, \, j, \, n+1/2}
\\
& \hspace{1.5em}
     - \bracketb{ V^{y} }_{i,   j, \, n+1/2} \, \bracketb{ B^{x} }_{i,   j, \, n+1/2}
     - \bracketb{ V^{x} }_{i,   j, \, n+1/2} \, \bracketb{ B^{y} }_{i,   j, \, n+1/2}
\Big] .
\end{align}
\end{subequations}
By slight abuse of notation $\psi^{x}_{i, \, j+1/2, \, n+1/2}$ denotes the average of $\psi^{x}_{i, \, j, \, n+1/2}$ and $\psi^{x}_{i, \, j+1, \, n+1/2}$ while $\psi^{y}_{i+1/2, \, j, \, n+1/2}$ denotes the average of $\psi^{y}_{i, \, j, \, n+1/2}$ and $\psi^{y}_{i+1, \, j, \, n+1/2}$ (c.f.~Equations~\eqref{eq:mhd_discrete_psi_action}).
Figure~\ref{fig:mhd_staggered_grid_scheme} shows the cells covered by the stencils of each equation.
The discretisation of the operators $\psi$ and $\phi$ is the very same as the one obtained by \citet{Gawlik:2011} and \citet{LiuWang:2001}, but the variational discretisation yields a different time stepping scheme, namely the implicit midpoint method, thus leading to improved conservation properties.
In particular, energy, magnetic helicity and cross helicity are preserved exactly (up to machine accuracy).

\subsection{Inertial Magnetohydrodynamics}

If we add effects due to electron inertia to the system, we have to consider the following additional terms in the action,
\begin{align}
\mcal{A} &= \int \Big[ ... + \sigma \cdot \big[ \obar{B} - B \big] - d_{e}^{2} \, \big[ \nabla \times \sigma \big] \cdot \big[ \nabla \times B \big] + ... \Big] \, dt \, dx \, dy .
\end{align}
The remaining parts are exactly the same as before, except that in some places $B$ is replaced with $\obar{B}$, which however does not affect the discretisation.
The first term is discretised in a similar way as the terms of the time derivatives, that is
\begin{subequations}
\begin{align}
\sigma^{x} \obar{B}^{x} \quad \rightarrow \quad
& \dfrac{1}{2} \Big[
  \sigma^{x}_{ i, \, j-1/2, \, n } \, \obar{B}^{x}_{i, \, j-1/2, \, n}
+ \sigma^{x}_{ i, \, j+1/2, \, n } \, \obar{B}^{x}_{i, \, j+1/2, \, n} 
  \Big] , \\
\sigma^{y} \obar{B}^{y} \quad \rightarrow \quad
& \dfrac{1}{2} \Big[
  \sigma^{y}_{ i-1/2, \, j, \, n } \, \obar{B}^{y}_{i-1/2, \, j, \, n}
+ \sigma^{y}_{ i+1/2, \, j, \, n } \, \obar{B}^{y}_{i+1/2, \, j, \, n} 
  \Big] .
\end{align}
\end{subequations}
The multiplier $\sigma$ is collocated at $n$, similar to $\gamma$ in the divergence-free constraint.
The curl terms, $\nabla \times \sigma$ and $\nabla \times B$, are discretised as
\begin{align}
\nabla \times B = \partial_{x} B^{y} - \partial_{y} B^{x}
\quad \rightarrow \quad
& \Big[
  ( \Delta_{x} \obar{B}^{y} )_{i, \, j, \, n}
- ( \Delta_{y} \obar{B}^{x} )_{i, \, j, \, n}
  \Big] ,
\end{align}
where the discrete derivatives $\Delta_{x}$ and $\Delta_{y}$ are defined as in \eqref{eq:discrete_field_derivatives_1}-\eqref{eq:discrete_field_derivatives_2}, but without the time average.
With this, the discrete Lagrangian for the electron inertia terms becomes
\begin{align}
L^{\text{E}}_{i,j,n}
\nonumber
&={} \dfrac{1}{2} \Big[
       \sigma^{x}_{ i, \, j-1/2, \, n } \, ( \obar{B}^{x}_{i, \, j-1/2, \, n} - B^{x}_{i, \, j-1/2, \, n} )
     + \sigma^{x}_{ i, \, j+1/2, \, n } \, ( \obar{B}^{x}_{i, \, j+1/2, \, n} - B^{x}_{i, \, j+1/2, \, n} )
   \Big] \\
\nonumber
&+{} \dfrac{1}{2} \Big[
       \sigma^{y}_{ i-1/2, \, j, \, n } \, ( \obar{B}^{y}_{i-1/2, \, j, \, n} - B^{y}_{i-1/2, \, j, \, n} )
     + \sigma^{y}_{ i+1/2, \, j, \, n } \, ( \obar{B}^{y}_{i+1/2, \, j, \, n} - B^{y}_{i+1/2, \, j, \, n} )
  \Big] \\
&-{} d_{e}^{2} \Big[
       ( \Delta_{x} \sigma^{y} )_{i, \, j, \, n}
     - ( \Delta_{y} \sigma^{x} )_{i, \, j, \, n}
  \Big]
  \Big[
       ( \Delta_{x} \obar{B}^{y} )_{i, \, j, \, n}
     - ( \Delta_{y} \obar{B}^{x} )_{i, \, j, \, n}
  \Big] .
\end{align}
Again, we can define an averaged Lagrangian, 
\begin{align}
L^{\text{E}}_{i,j,n+1/2} = \tfrac{1}{2} \Big[ L^{\text{D}}_{i,j,n} + L^{\text{D}}_{i,j,n+1} \Big] ,
\end{align}
so that all Lagrangians are collocated at the same point in space and time.
Requiring stationarity of the modified discrete action, 
\begin{align}
\delta \obar{\mcal{A}}_{d} [\obar{\phy}_{d}] = \delta \sum \limits_{i=1}^{n_{x}} \sum \limits_{j=1}^{n_{y}} \sum \limits_{n=0}^{n_{t}-1} \Big[ L^{\text{M}}_{i,j,n+1/2} + L^{\text{D}}_{i,j,n+1/2} + L^{\text{I}}_{i,j,n+1/2} + L^{\text{E}}_{i,j,n+1/2} \Big] = 0 , 
\end{align}
yields the discrete inertial MHD equations,
\begingroup
\allowdisplaybreaks
\begin{subequations}\label{eq:variational_integrator_inertial_mhd}
\begin{align}
\nonumber
& ( \Delta_{t} V^{x} )_{i, \, j+1/2, \, n+1/2}
+ \psi^{x}_{i, \, j+1/2, \, n+1/2} (V,V) \\
& \hspace{8em}
- \psi^{x}_{i, \, j+1/2, \, n+1/2} (\obar{B},B)
+ ( \Delta_{x} P )_{i, \, j+1/2, \, n+1/2} = 0 ,
\\
\nonumber
& ( \Delta_{t} V^{y} )_{i+1/2, \, j, \, n+1/2}
+ \psi^{y}_{i+1/2, \, j, \, n+1/2} (V,V) \\
& \hspace{8em}
- \psi^{y}_{i+1/2, \, j, \, n+1/2} (\obar{B},B)
+ ( \Delta_{y} P )_{i+1/2, \, j, \, n+1/2} = 0 ,
\\
& ( \Delta_{t} \obar{B} )^{x}_{i, \, j+1/2, \, n+1/2}
+ \phi^{x}_{i, \, j+1/2, \, n+1/2} (V,\obar{B}) = 0 ,
\\
& ( \Delta_{t} \obar{B} )^{y}_{i+1/2, \, j, \, n+1/2}
+ \phi^{y}_{i+1/2, \, j, \, n+1/2} (V,\obar{B}) = 0 ,
\\
&  ( \Delta_{x} V^{x} )_{i+1/2, \, j+1/2, \, n+1/2}
 + ( \Delta_{y} V^{y} )_{i+1/2, \, j+1/2, \, n+1/2} = 0 ,
\\
& B^{x}_{i, \, j+1/2, \, n} + d_{e}^{2} \big( ( \Delta_{x} \Delta_{y} B^{y} )_{i, \, j+1/2, \, n} - ( \Delta_{y} \Delta_{y} B^{x} )_{i, \, j+1/2, \, n} \big) = \obar{B}^{x}_{i, \, j+1/2, \, n} ,
\\
& B^{y}_{i+1/2, \, j, \, n} + d_{e}^{2} \big( ( \Delta_{y} \Delta_{x} B^{x} )_{i+1/2, \, j, \, n} - ( \Delta_{x} \Delta_{x} B^{y} )_{i+1/2, \, j, \, n} \big) = \obar{B}^{y}_{i+1/2, \, j, \, n} ,
\end{align}
\end{subequations}
\endgroup
with $\obar{\phy}_{d}$ denoting the discrete solution,
\begin{align}
\obar{\phy}_{d} = \Big\{
\nonumber
& V^{x}_{i,j+1/2,n}, \, V^{y}_{i+1/2,j,n}, \, \obar{B}^{x}_{i,j+1/2,n}, \, \obar{B}^{y}_{i+1/2,j,n}, \, P_{i+1/2,j+1/2,m+1/2}, \, B^{x}_{i,j+1/2,n}, \, B^{y}_{i+1/2,j,n}, \\
\nonumber
& \alpha^{x}_{i,j+1/2,m+1/2}, \, \alpha^{y}_{i+1/2,j,m+1/2}, \, \beta^{x}_{i,j+1/2,m+1/2}, \, \beta^{y}_{i+1/2,j,m+1/2}, \, \gamma_{i+1/2,j+1/2,n}, \, \sigma_{i+1/2,j+1/2,n} \\
& \Big\vert \; 1 \leq i \leq n_{x} , \, 1 \leq j \leq n_{y} , \, 0 \leq m < n_{t} - 1 , \, 0 \leq n \leq n_{t} \Big\} ,
\end{align}
and the discrete derivatives being defined by
\begin{subequations}
\begin{align}
\nonumber
( \Delta_{x} \Delta_{y} B^{x} )_{i+1/2, \, j, \, n} &= \dfrac{1}{h_{x} h_{y}} \Big[ 
     \big( B^{x}_{i+1, \, j+1/2, \, n} - B^{x}_{i+1, \, j-1/2, \, n} \big) \\
& \hspace{8em}
   - \big( B^{x}_{i, \, j+1/2, \, n} - B^{x}_{i, \, j-1/2, \, n} \big)
\Big] , \\
\nonumber
( \Delta_{x} \Delta_{y} B^{y} )_{i, \, j+1/2, \, n} &= \dfrac{1}{h_{x} h_{y}} \Big[
     \big( B^{y}_{i+1/2, \, j+1, \, n} - B^{y}_{i-1/2, \, j+1, \, n} \big) \\
& \hspace{8em}
   - \big( B^{y}_{i+1/2, \, j, \, n} - B^{y}_{i-1/2, \, j, \, n} \big)
\Big] , \\
( \Delta_{x} \Delta_{x} B^{y} )_{i+1/2, \, j, \, n} &= \dfrac{1}{h_{x}^{2}} \Big[ B^{y}_{i+3/2, \, j, \, n} - 2 \, B^{y}_{i+1/2, \, j, \, n} + B^{y}_{i-1/2, \, j, \, n} \Big] , \\
( \Delta_{y} \Delta_{y} B^{x} )_{i, \, j+1/2, \, n} &= \dfrac{1}{h_{y}^{2}} \Big[ B^{x}_{i, \, j+3/2, \, n} - 2 \, B^{x}_{i, \, j+1/2, \, n} + B^{x}_{i, \, j-1/2, \, n} \Big] ,
\end{align}
\end{subequations}
thus using only points within the same red and green grid cells in Figure~\ref{fig:mhd_staggered_grid_scheme} as the other discrete operators.

\section{Numerical Examples}\label{sec:examples}

In this section, we use the variational integrator for the inertial MHD system to simulate a current sheet, first in the ideal case, in order to verify that the solution is free from spurious reconnection, and then in the inertial case, where reconnection is supposed to happen. We compare our results with another variational integrator for reduced MHD~\cite{KrausTassiGrasso:2016}.
Numerical examples for the ideal MHD integrator have already been reported in~\citet{KrausMaj:2017}.

The variational integrator~(\ref{eq:variational_integrator_ideal_mhd}) is implemented using Python~\cite{ScopatzHuff:2015, Langtangen:2014}, Cython~\cite{Behnel:2010}, PETSc~\cite{petsc-web-page, petsc-user-ref} and petsc4py~\cite{Dalcin:2011}. Visualisation was done using NumPy~\cite{vanDerWalt:2011}, SciPy~\cite{scipy-web-page} and matplotlib~\cite{Hunter:2007}.
The nonlinear system is solved with Picard's method, where in each iteration three linear systems are solved: the momentum equation~\eqref{eq:inertial_mhd_equations_1} and the divergence constraint~\eqref{eq:inertial_mhd_equations_3} for $V$ and $P$, the induction equation~\eqref{eq:inertial_mhd_equations_2} for $\obar{B}$, and the constraint~\eqref{eq:inertial_mhd_equations_4} for $B$.
Each linear system is solved via LU decomposition with SuperLU~\cite{Li:2005, superlu-web-page}.
More efficient solvers can be constructed using iterative solvers like GMRES and conjugate gradients with appropriate preconditioning for the various sub-systems, but this is not the topic of this work.
The tolerance of the nonlinear solver is set to $10^{-10}$ or smaller, which for the time steps chosen is usually reached after $3-5$ iterations.

\subsection{Diagnostics}

In the following we give discrete expressions of the conserved quantities, energy~(\ref{eq:inertial_mhd_energy}), cross helicity~(\ref{eq:inertial_mhd_cross_helicity}) and magnetic helicity~(\ref{eq:inertial_mhd_magnetic_helicity}), which are monitored in the simulations, as well as the discrete equations for the reconstruction of the vector potential and the current density.

\subsubsection*{Energy}

The total energy of the system is the sum of kinetic energy and magnetic energy, which are computed by
\begin{align}
E_{\text{kin}} (t_n) &= h_{x} \, h_{y} \, \dfrac{1}{2} \sum \limits_{i, \, j} \Big[
    \big( V^{x}_{i, \, j+1/2, \, n} \big)^{2} + \big( V^{y}_{i+1/2, \, j, \, n}  \big)^{2} \Big] , \\
E_{\text{mag}} (t_n) &= h_{x} \, h_{y} \, \dfrac{1}{2} \sum \limits_{i, \, j} \Big[ B^{x}_{i, \, j+1/2, \, n} \, \obar{B}^{x}_{i, \, j+1/2, \, n} + B^{y}_{i+1/2, \, j, \, n} \, \obar{B}^{y}_{i+1/2, \, j, \, n} \Big] .
\end{align}
As there is no dissipation term in the ideal and inertial MHD equations, \eqref{eq:ideal_mhd_equations} and \eqref{eq:inertial_mhd_equations}, respectively, the total energy should always be preserved.

\subsubsection*{Cross Helicity}

The cross helicity is the integral of the scalar product of the velocity and magnetic field,
\begin{align}
C_{\mrm{CH}} (t_n)
&= h_{x} \, h_{y} \, \dfrac{1}{2} \sum \limits_{i, \, j} \Big[
   V^{x}_{i, \, j+1/2, \, n} \, \obar{B}^{x}_{i, \, j+1/2, \, n}
 + V^{y}_{i+1/2, \, j, \, n} \, \obar{B}^{y}_{i+1/2, \, j, \, n} \Big] .
\end{align}
In ideal and inertial MHD, the parallel components of the velocity and magnetic field do not interact, so that the integral of their product over the spatial domain stays constant.

\subsubsection*{Magnetic Helicity}

In two dimensions, magnetic helicity reduces to the integral of the vector potential,
\begin{align}
C_{\mrm{MH}} (t_n) &= h_{x} \, h_{y} \, \sum \limits_{i, \, j} \obar{A}_{i, \, j, \, n} ,
\end{align}
where $\obar{A}_{i,j}$ is reconstructed as described below.

\subsubsection*{Vector Potential}

In two dimensions, the magnetic field is given by
\begin{align}
\obar{B}^{x} = \partial_{y} \obar{A}
\hspace{3em}
\text{and}
\hspace{3em}
\obar{B}^{y} = - \partial_{x} \obar{A} ,
\end{align}
where $\obar{A}$ is the $z$-component of the generalised vector potential.
The potential $\obar{A}$ is collocated at the cell centres of Figure~\subref*{fig:mhd_staggered_grid_momentum}.
Therefore these equations are discretised as
\begin{align}\label{eq:diagnostics_B}
\obar{B}^{x}_{i, \, j+1/2, \, n} &=   \dfrac{\obar{A}_{i, \, j+1, \, n} - \obar{A}_{i, \, j, \, n}}{h_{y}}   &
& \text{and} &
\obar{B}^{y}_{i+1/2, \, j, \, n} &= - \dfrac{\obar{A}_{i+1, \, j, \, n} - \obar{A}_{i, \, j, \, n}}{h_{x}} .
\end{align}
Equations \eqref{eq:diagnostics_B} can be rewritten as recurrence relations for $\obar{A}_{i,j}$, namely
\begin{align}
\obar{A}_{i, \, j+1, \, n} &= \obar{A}_{i, \, j, \, n} + h_{y} \, \obar{B}^{x}_{i, \, j+1/2, \, n} &
& \text{and} &
\obar{A}_{i+1, \, j, \, n} &= \obar{A}_{i, \, j, \, n} - h_{x} \, \obar{B}^{y}_{i+1/2, \, j, \, n} .
\end{align}
The vector potential is obtained by fixing the value of $\obar{A}$ in the point $(i,j)=(1,1)$ and looping over the whole grid, using the first equation to compute columns and the second to jump between rows, or the other way around.
To which value $\obar{A}_{1,1}$ is fixed is not important as $\obar{A}$ is determined only up to a constant.
The contour lines of the generalised magnetic potential $\obar{A}$ correspond to field lines of the generalised magnetic field $\obar{B}$. Hence, $\obar{A}$ is an important diagnostic.

\subsection{Current Sheet}
\label{sec:current_sheet}

\begin{figure}[th]
\centering
\subfloat[t=0]{
\includegraphics[width=.32\textwidth]{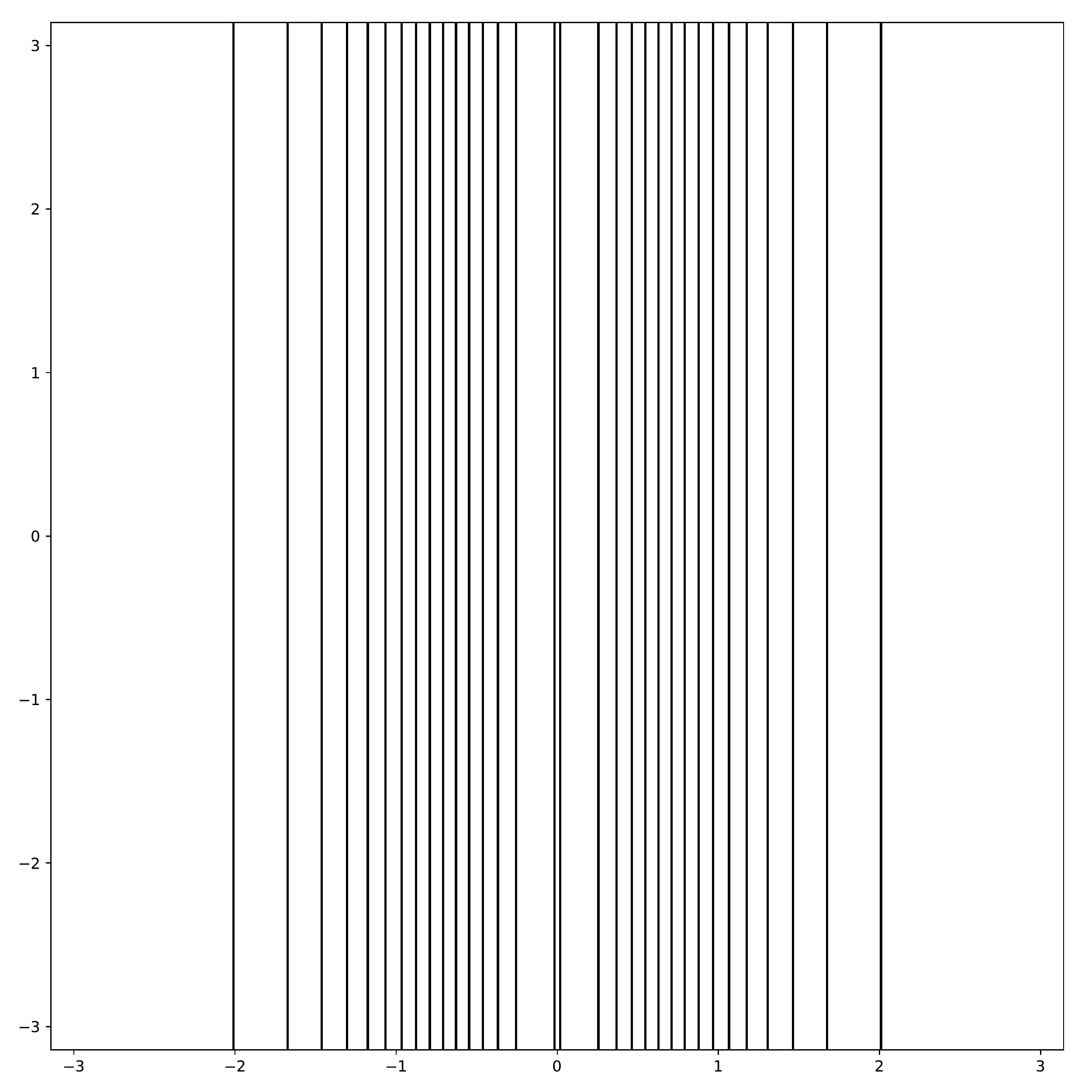}
}
\subfloat[t=50]{
\includegraphics[width=.32\textwidth]{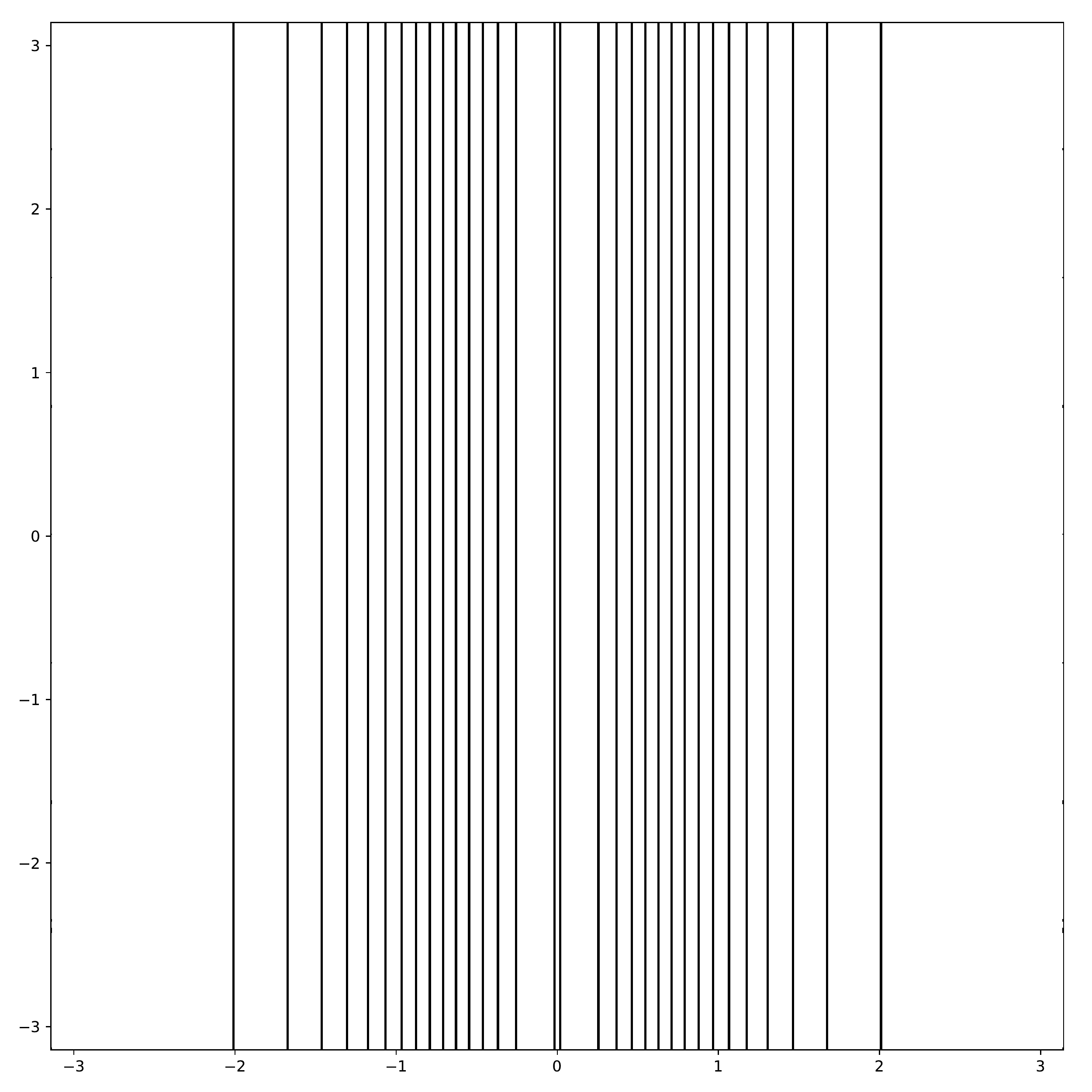}
}
\subfloat[t=100]{
\includegraphics[width=.32\textwidth]{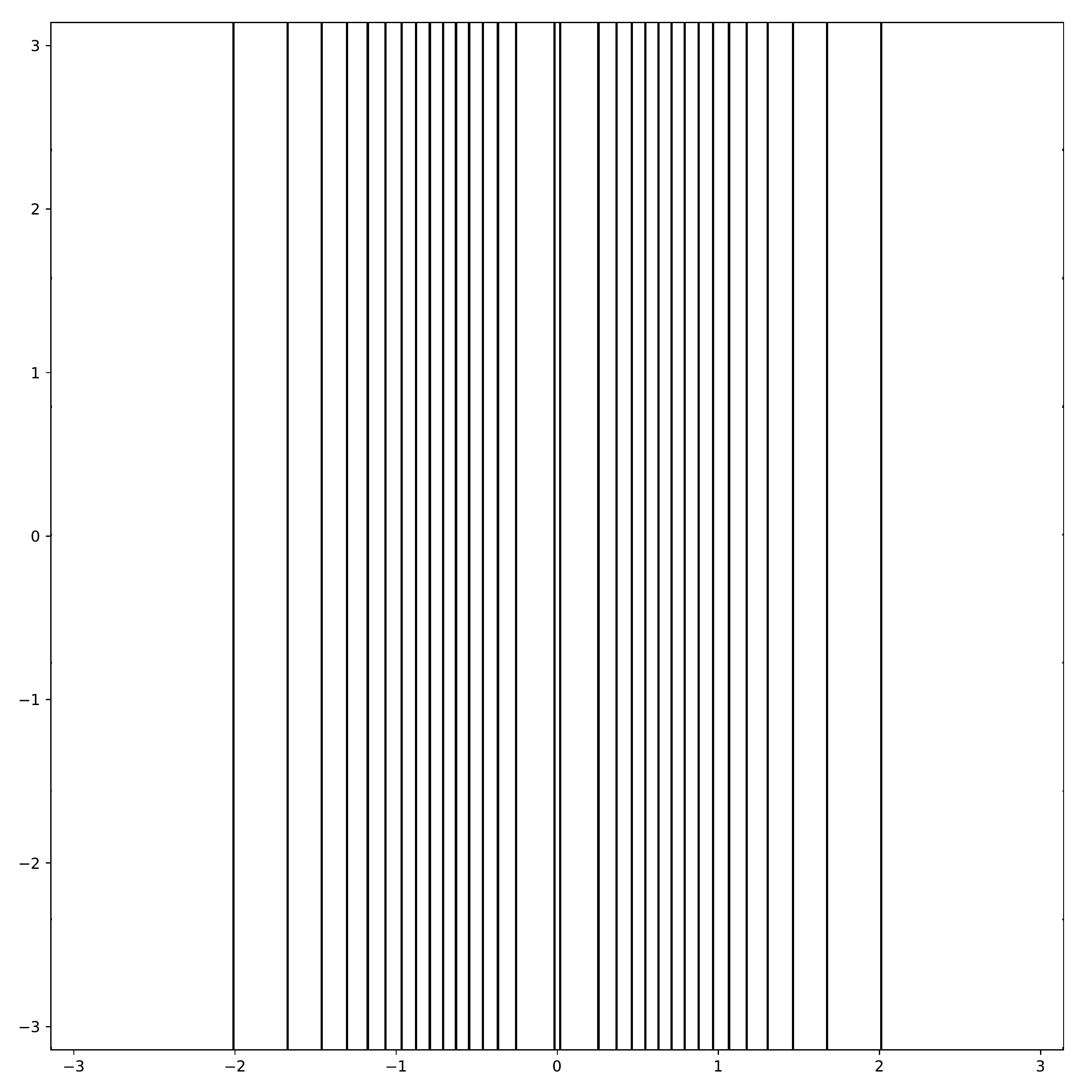}
}

\caption{Current sheet. Contour lines of the flux function $A$.}
\label{fig:mhd_current_sheet_psi}
\end{figure}

\begin{figure}[th]
\centering
\includegraphics[width=.45\textwidth]{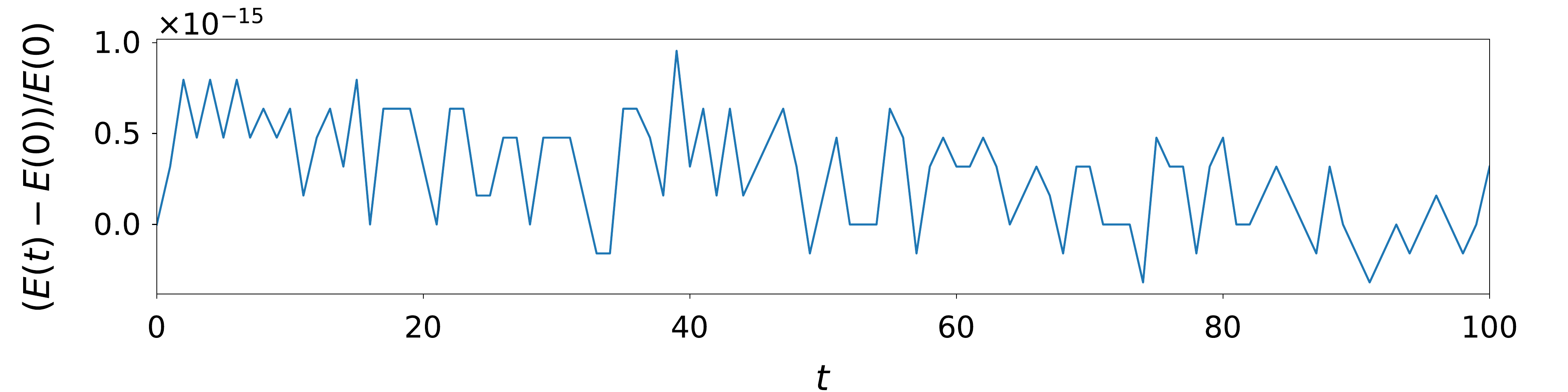}

\includegraphics[width=.45\textwidth]{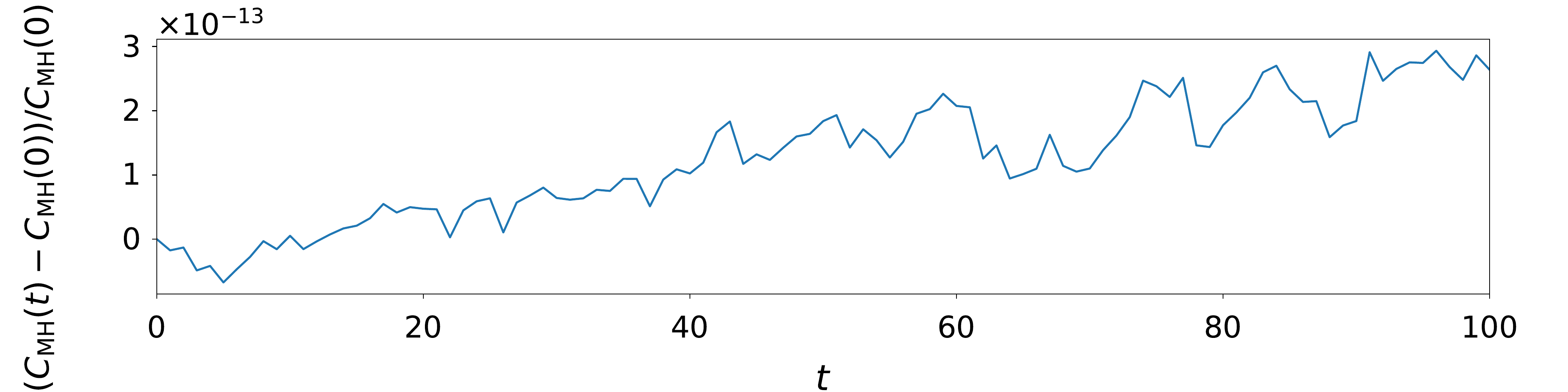}

\includegraphics[width=.45\textwidth]{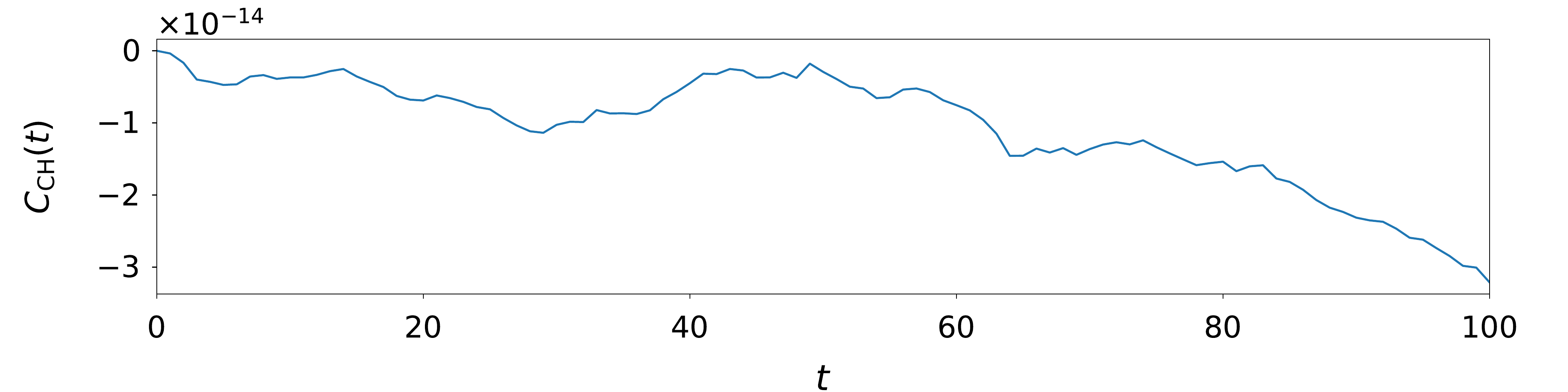}

\caption{Current sheet with the variational integrator. Error of the total energy $E$, magnetic helicity $C_{\mathrm{MH}}$ and cross helicity $C_{\mathrm{CH}}$.}
\label{fig:mhd_current_sheet_timetraces}
\end{figure}

In this example we adopt the variational integrator~\eqref{eq:variational_integrator_ideal_mhd} to solve the ideal MHD equations~\eqref{eq:ideal_mhd_equations}. We consider the following initial condition
\begin{align}\label{eq:current_sheet_initcondcs}
\phi &= \phi_{0} \, \big( \cos (x+y) - \cos (x-y) \big) , &
A &= \dfrac{A_{0}}{\cosh^{2} (x)} ,
\end{align}
with $A_{0} = 1.29$ and $\phi_{0} = 10^{-3}$, which in the field description used here corresponds to
\begin{align}\label{eq:current_sheet_initcondcs2}
V &= \phi_{0} \begin{pmatrix}
\sin (x+y) + \sin (x-y) \\
\sin (x-y) - \sin (x+y) \\
\end{pmatrix} , &
B &= A_{0} \begin{pmatrix}
0 \\
- 2 \tanh (x) / \cosh(x)^2 \\
\end{pmatrix} .
\end{align}
The initial pressure is set to $p = 1$ everywhere.
These initial conditions lead to the formation of a current sheet centred at $x=0$. The same initial condition was adopted for collisionless reconnection studies~\cite{Grasso:2006, TassiGrasso:2010}.
The spatial domain is $(x,y) \in [-\pi , +\pi) \times [-\pi , +\pi)$ with periodic boundaries, resolved by $n_{x} \times n_{y} = 1024 \times 512$ grid points.
In order to satisfy the periodicity condition in the $x$ direction a Fourier series representation of the equilibrium magnetic field has been adopted. Namely, we expand the expression for $B$ in Equation~\eqref{eq:current_sheet_initcondcs2} in a Fourier series, truncated up to $22$ modes. This truncation has already been shown~\cite{Grasso:2006} to provide a good representation of the equilibrium flux function. 

Figure~\ref{fig:mhd_current_sheet_psi} shows the magnetic potential $A$ at various points in time.
The topology of the contour lines of the magnetic potential is preserved throughout the whole simulation.
Artificial reconnection due to spurious effects of the numerics is absent.
This is also reflected in the good conservation properties regarding energy, magnetic helicity and cross helicity (see Figure~\ref{fig:mhd_current_sheet_timetraces}).

\subsection{Magnetic Reconnection}

In the previous example we verified that in the ideal case, the variational integrator~\eqref{eq:variational_integrator_ideal_mhd} for the ideal MHD system~\eqref{eq:ideal_mhd_equations} is free of artificial reconnection due to numerical resistivity or other spurious effects.
Now we consider the variational integrator~\eqref{eq:variational_integrator_inertial_mhd} for the inertial MHD system~\eqref{eq:inertial_mhd_equations}.
We use the same setup and the same initial condition as in the example of Section~\ref{sec:current_sheet}, but we solve the inertial MHD model with electron inertia effects, setting the electron skin depth $d_e$ to $0.2$, so that reconnection of magnetic field lines is expected to take place. We compare the results obtained from the variational integrator for ideal MHD with those obtained from a variational integrator for reduced MHD~\cite{KrausTassiGrasso:2016}.
The simulations have been performed on a spatial domain of $(x,y) \in [-\pi , +\pi) \times [-\pi , +\pi)$ with periodic boundaries using $n_{x} \times n_{y} = 1024 \times 512$ grid points.
In both simulations we use time steps $h_{t} = 0.1$ for $t \in [0, 22]$ and $h_{t} = 0.01$ for $t \in [22,30]$. For the ideal MHD solver the time step needs to be reduced further to $h_{t} = 0.001$ during the strongly nonlinear phase, $t \in [30,32]$, while for the reduced MHD solver we also use $h_{t} = 0.01$ for $t \in [30,32]$.
The necessity for reducing the time step towards the end of the simulation with the ideal MHD integrator stems fron the strong nonlinearity in conjunction with using Picard's method for solving the nonlinear system. It is expected that with Newton's method we can use the same time step as with the reduced MHD integrator, for which we are indeed using Newton's method to solve the nonlinear system.

\begin{figure}[t]
	\centering
	\includegraphics[width=.7\textwidth]{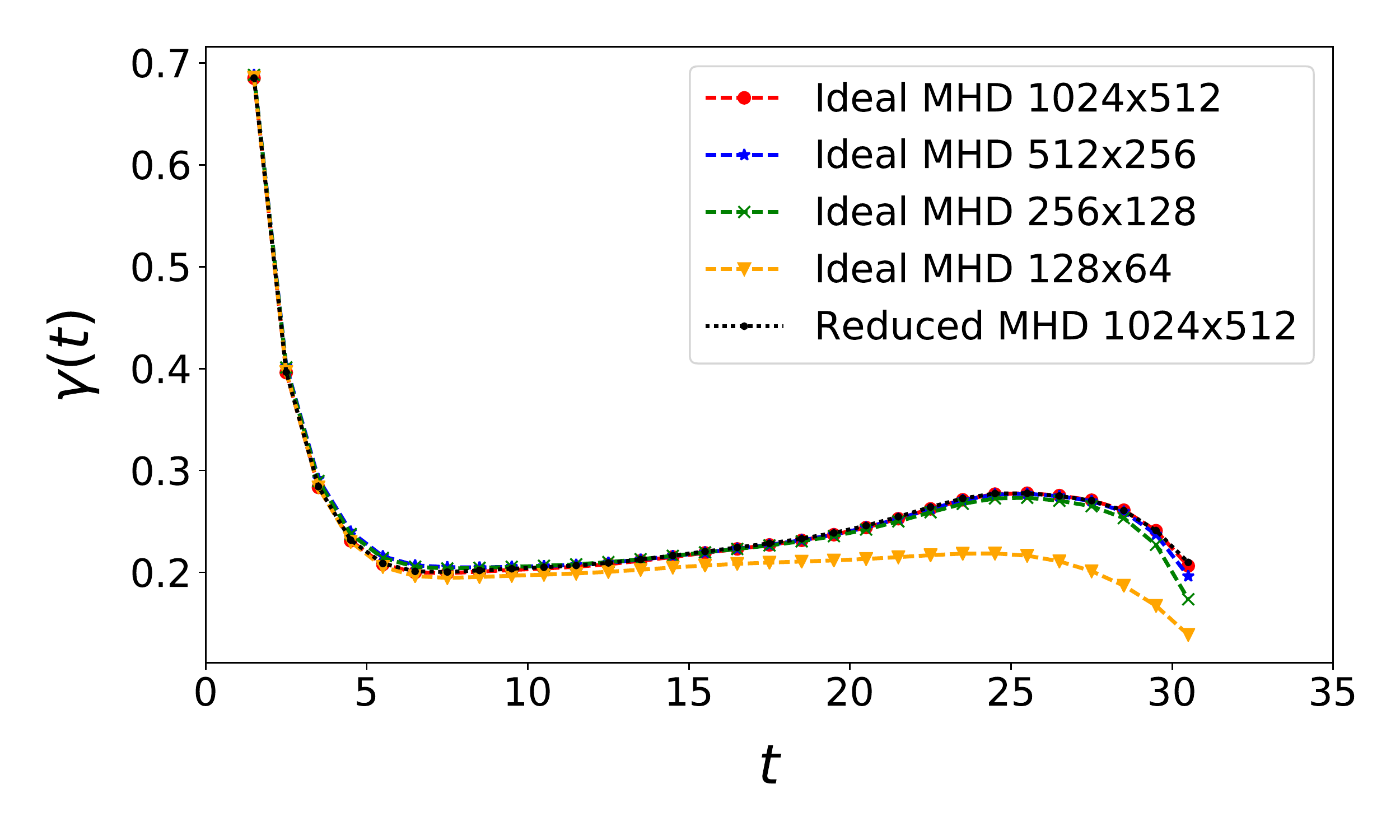}
	\caption{Growth rate $\gamma$ of the magnetic island computed with both integrators.}
	\label{fig:reconnection_growth_rate}
\end{figure}

An important quantity to consider in reconnection studies is the linear growth rate of the initial perturbation, which is defined as
\begin{align}
\gamma(t) = \frac{d}{dt} \ln (A(t,0,\pi)-A(0,\pi,0)) .
\end{align}
From Figure~\ref{fig:reconnection_growth_rate} one sees that the growth rate follows the expected behaviour consisting of a transient phase up to about $t=6$, followed by the linear phase, from about $t=6$ to $t=12$, where $\gamma$ is nearly constant, before entering the nonlinear phase for $t>12$. We observe that growth rates determined with the two variational integrators are almost identical.

\begin{figure}[p]
\centering
\subfloat[t=28]{
\includegraphics[width=.42\textwidth]{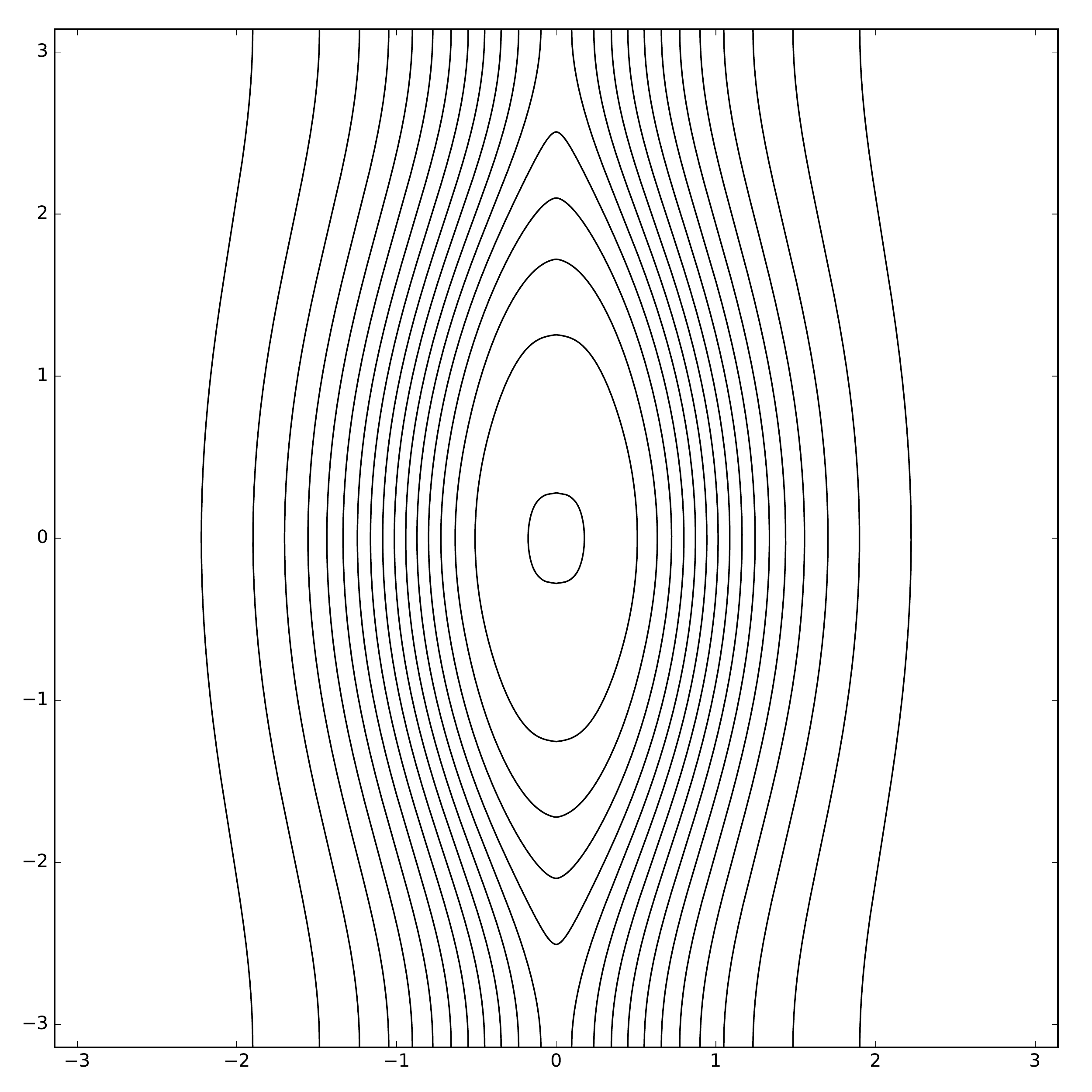}
}
\subfloat[t=28]{
\includegraphics[width=.42\textwidth]{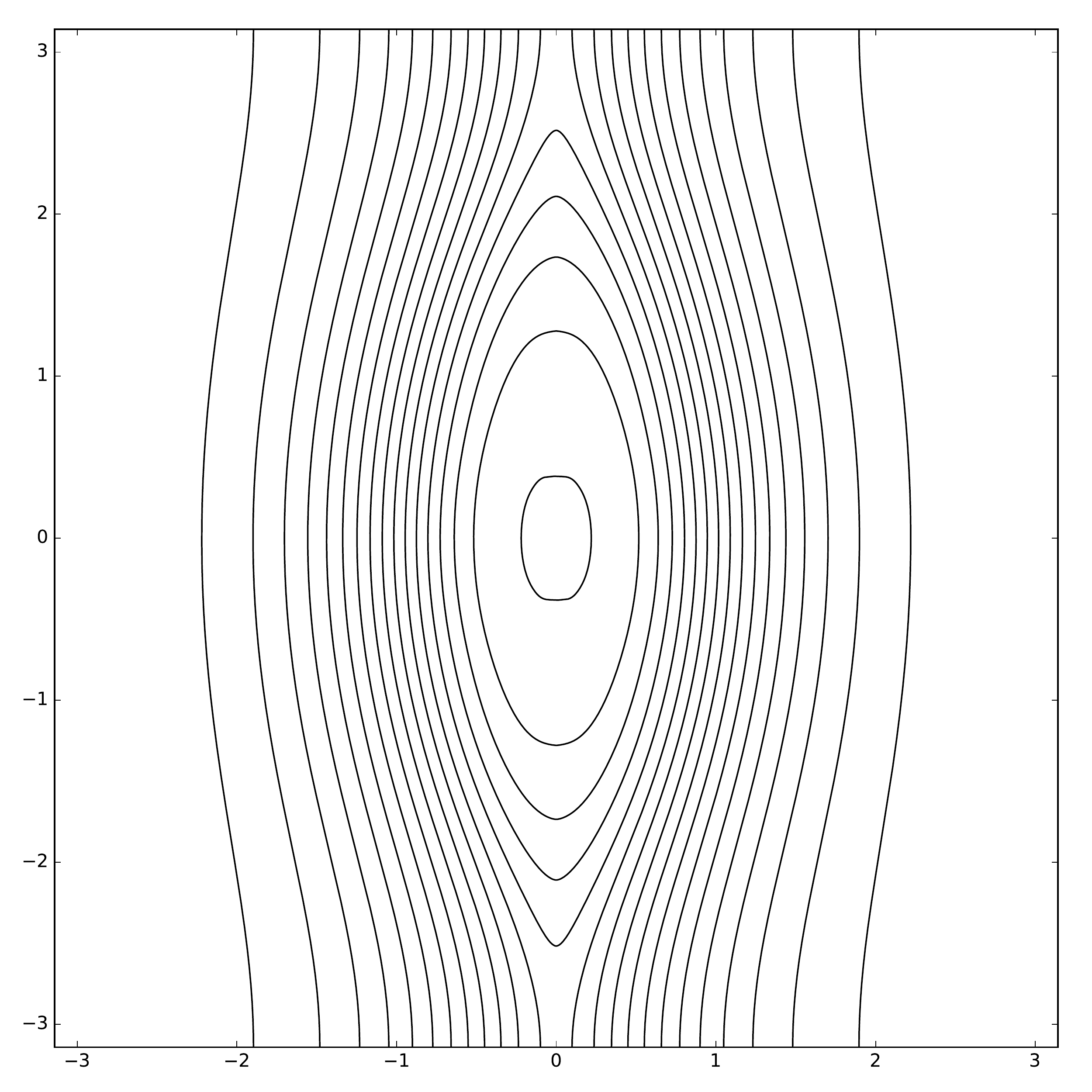}
}

\subfloat[t=30]{
\includegraphics[width=.42\textwidth]{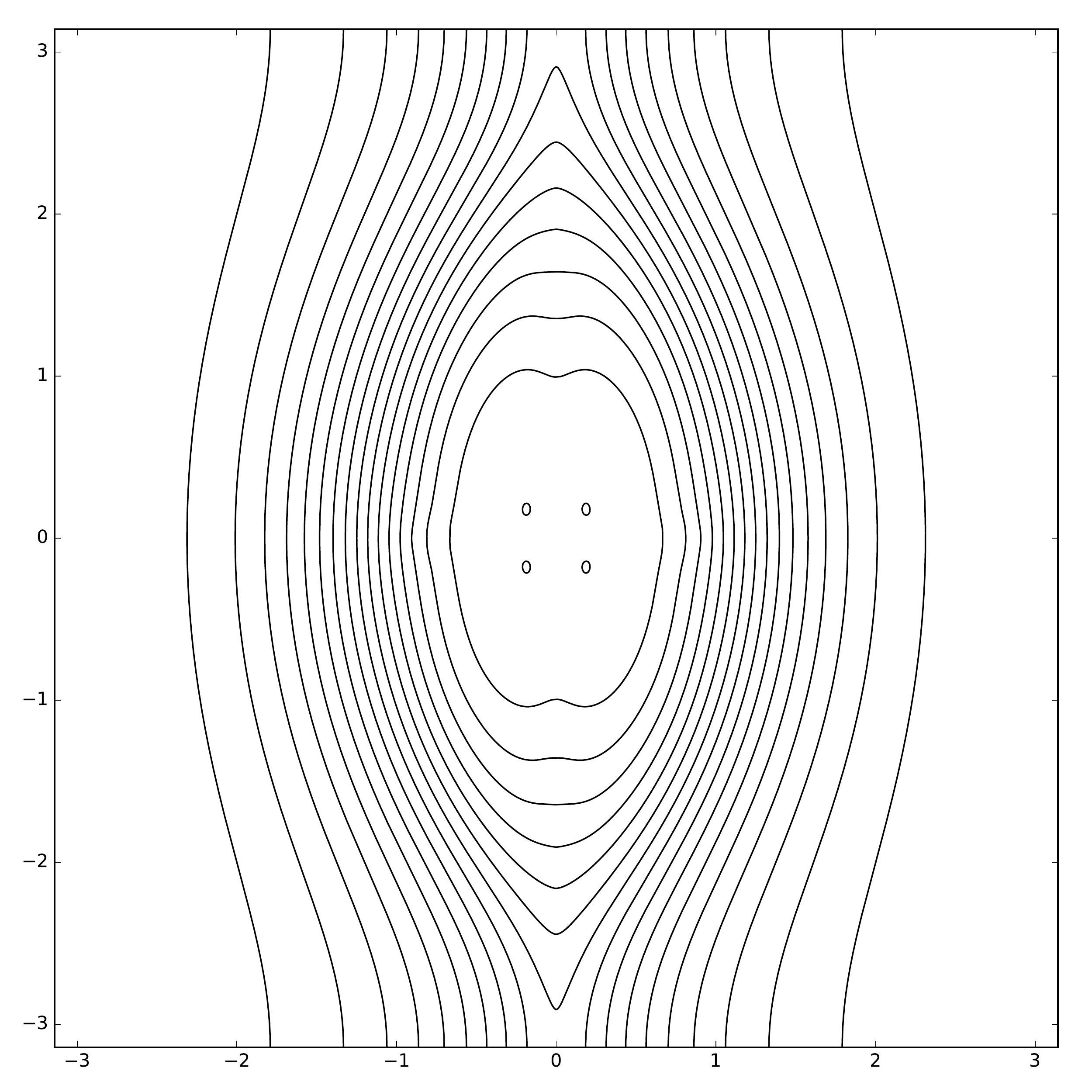}
}
\subfloat[t=30]{
\includegraphics[width=.42\textwidth]{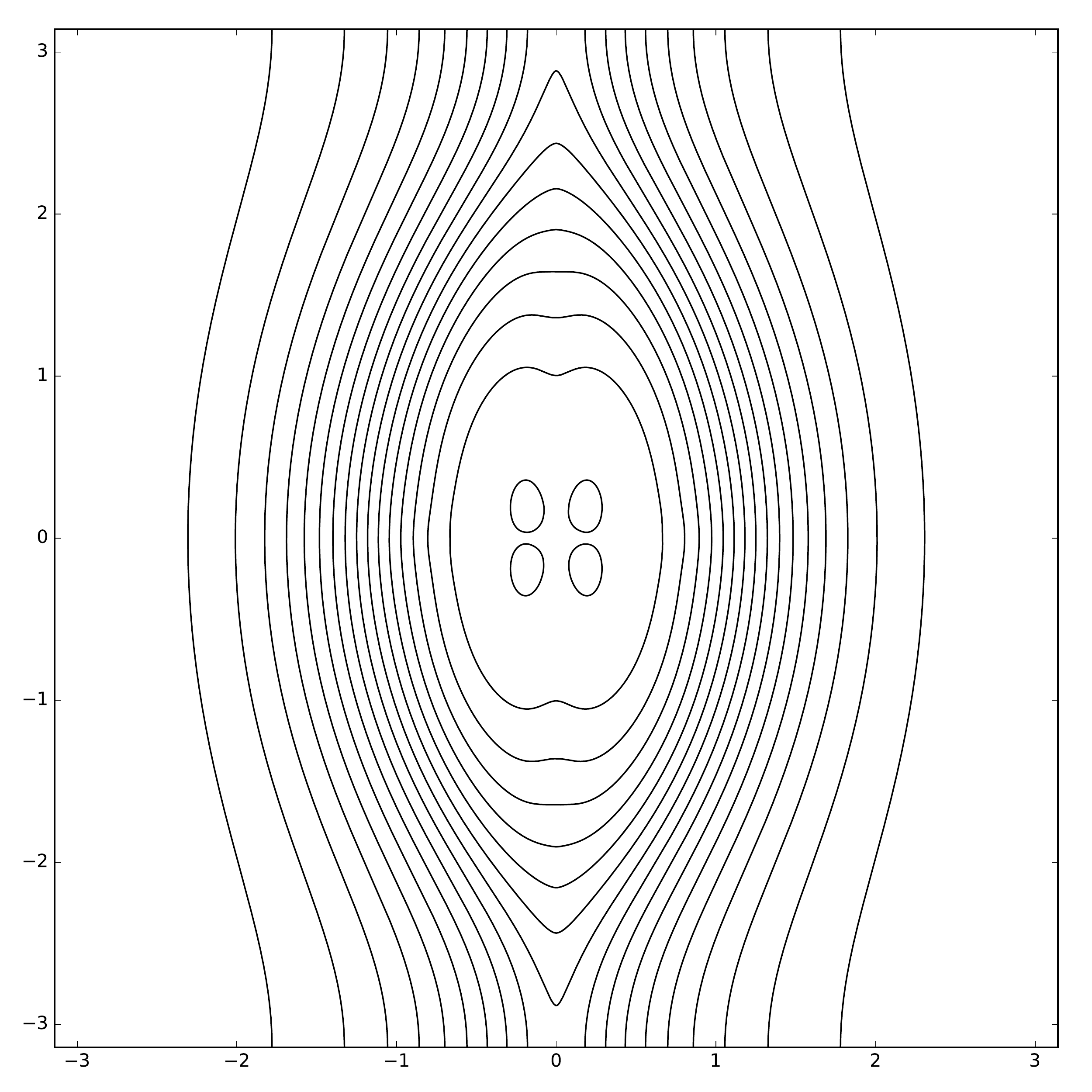}
}

\subfloat[t=32]{
\includegraphics[width=.42\textwidth]{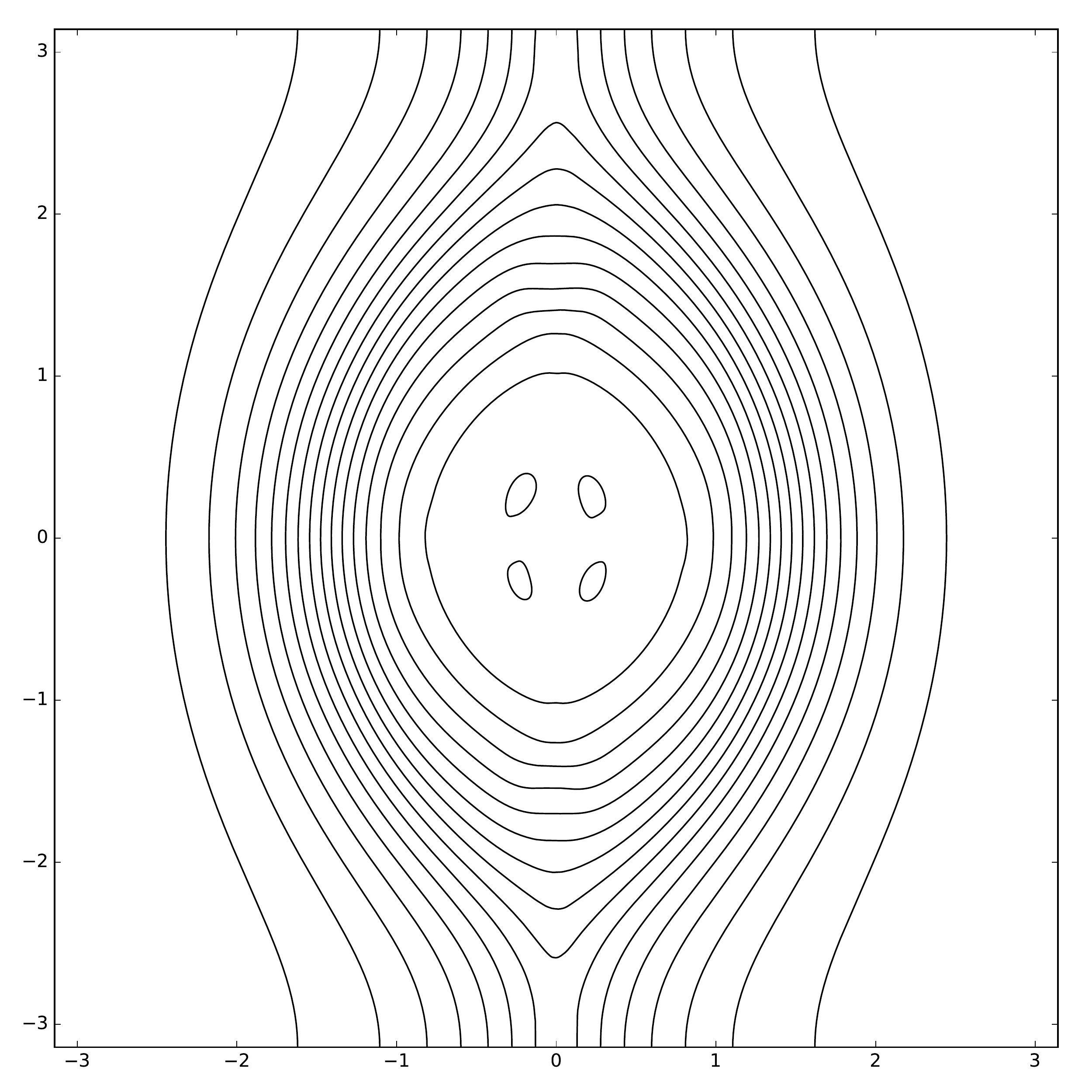}
}
\subfloat[t=32]{
\includegraphics[width=.42\textwidth]{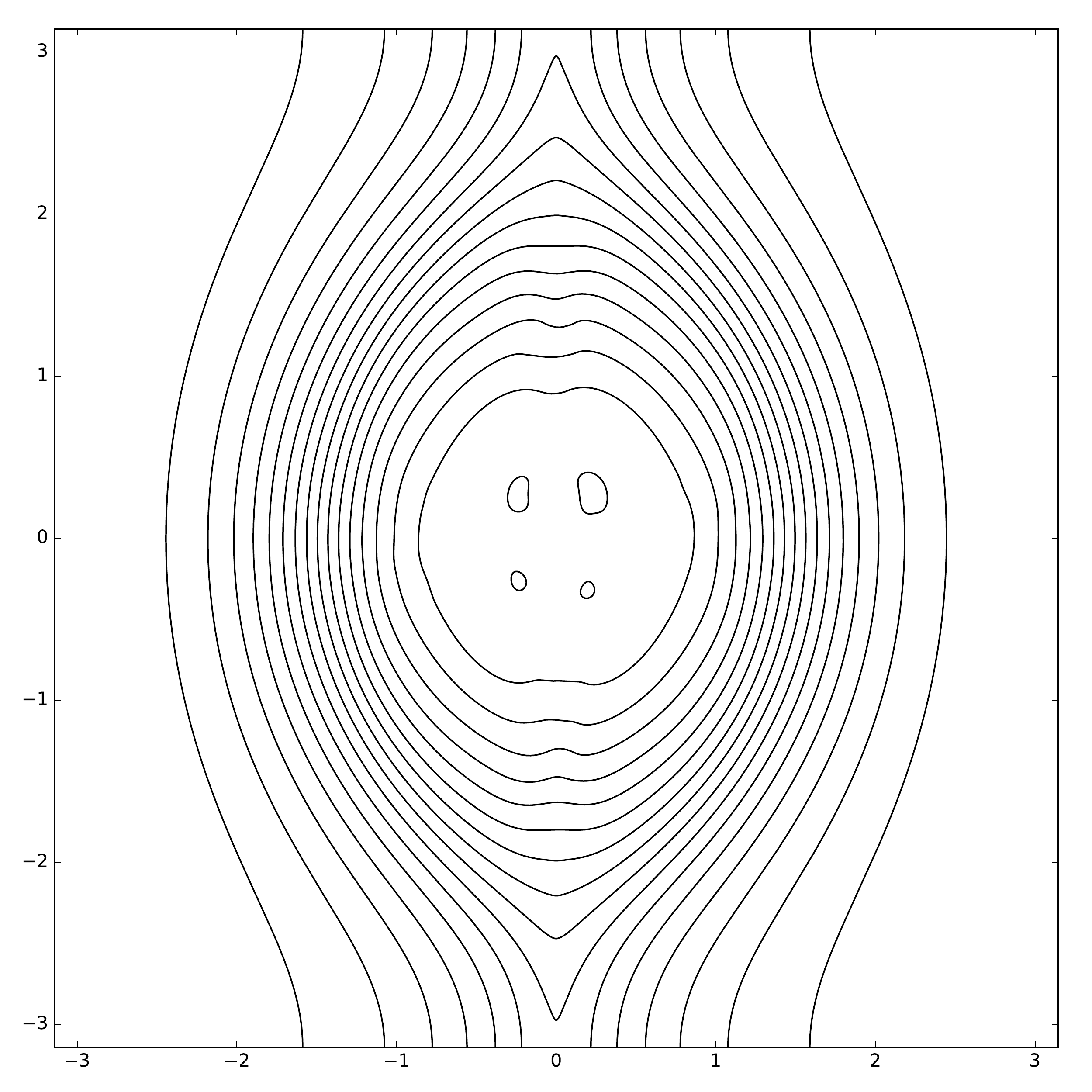}
}

\caption{Magnetic reconnection with ideal MHD variational integrator (left) and reduced MHD variational integrator (right). Vector potential $A$.}
\label{fig:mhd_reconnection_psi}
\end{figure}

\begin{figure}[p]
\centering
\subfloat[t=28]{
\includegraphics[width=.42\textwidth]{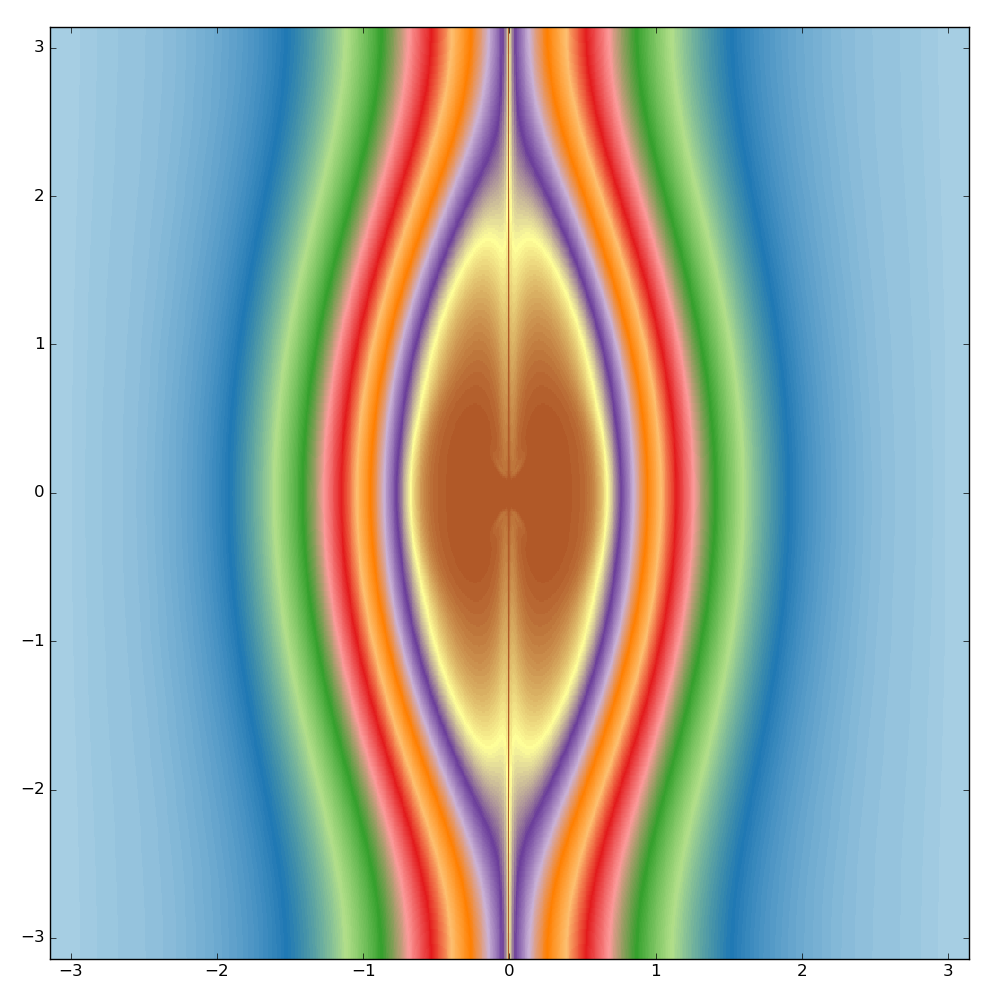}
}
\subfloat[t=28]{
\includegraphics[width=.42\textwidth]{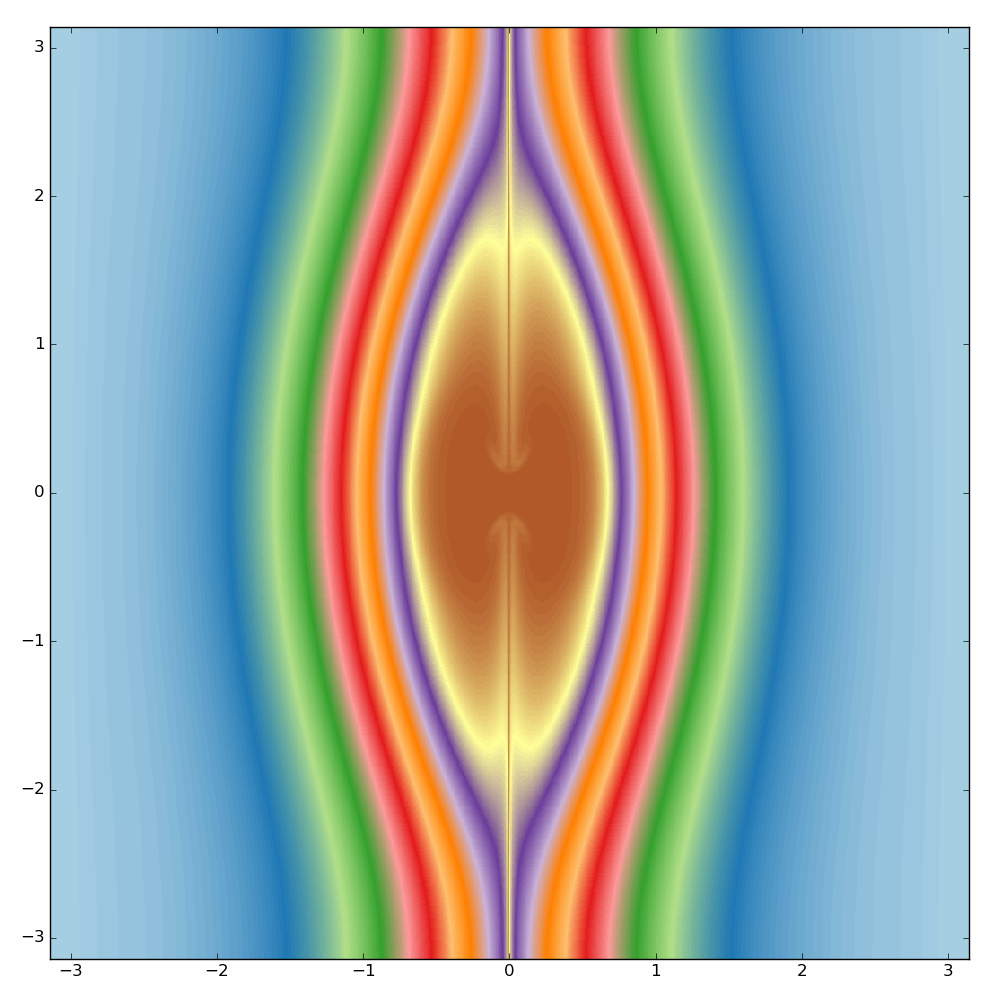}
}

\subfloat[t=30]{
\includegraphics[width=.42\textwidth]{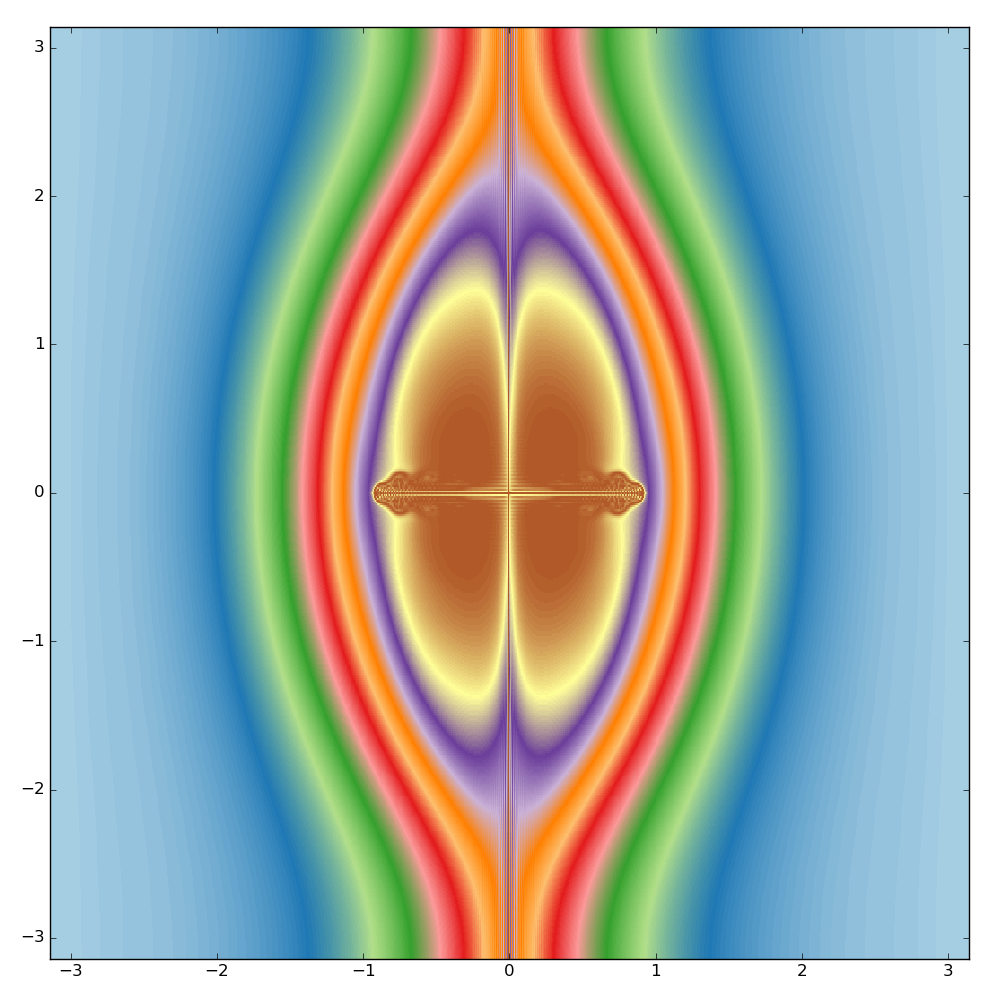}
}
\subfloat[t=30]{
\includegraphics[width=.42\textwidth]{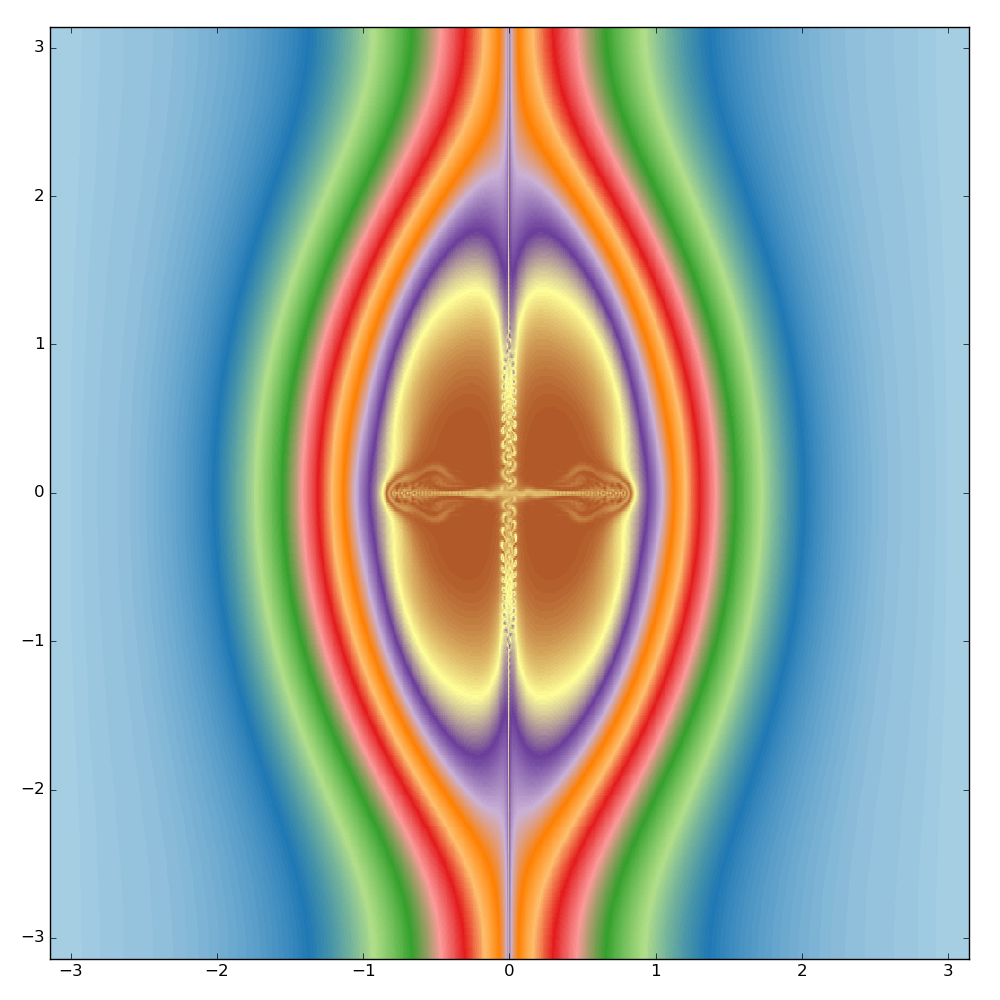}
}

\subfloat[t=32]{
\includegraphics[width=.42\textwidth]{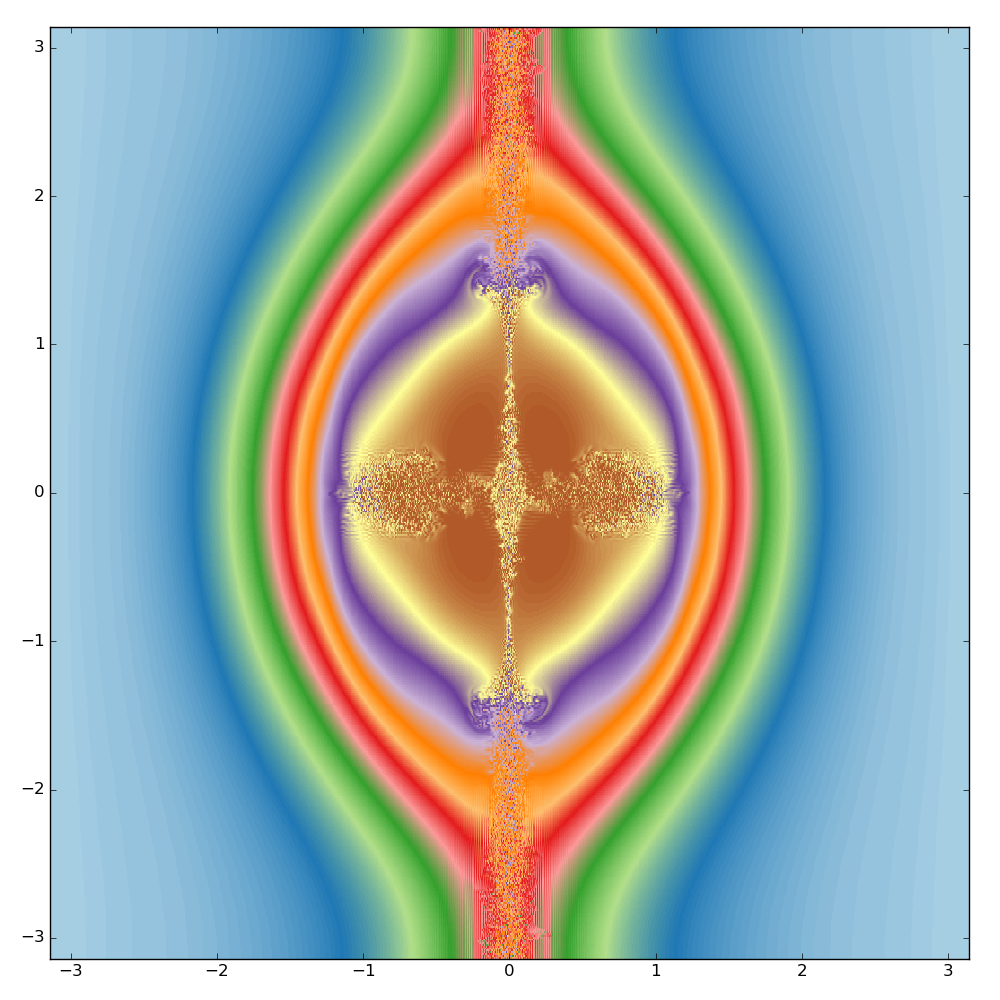}
}
\subfloat[t=32]{
\includegraphics[width=.42\textwidth]{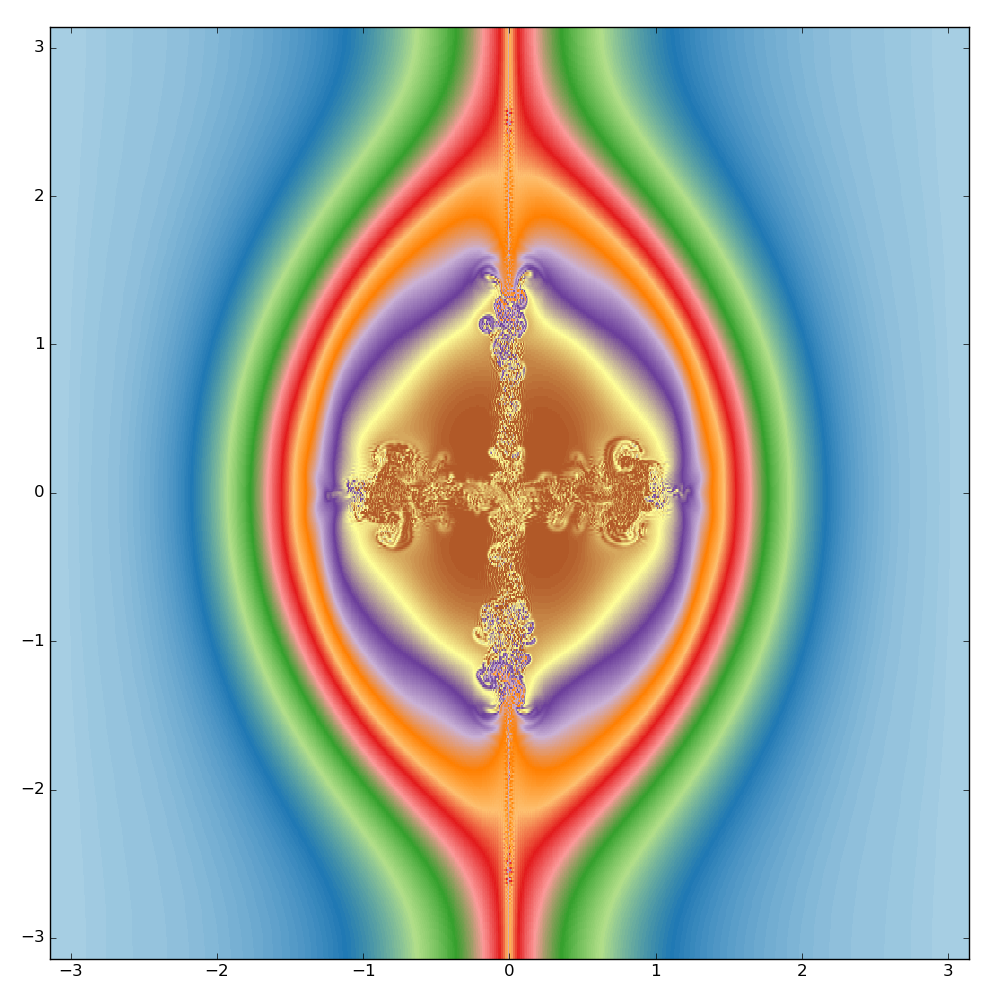}
}

\caption{Magnetic reconnection with ideal MHD variational integrator (left) and reduced MHD variational integrator (right). Generalised vector potential $\obar{A}$. Fixed colour scale.}
\label{fig:mhd_reconnection_psie}
\end{figure}

\begin{figure}[p]
\centering
\includegraphics[width=.45\textwidth]{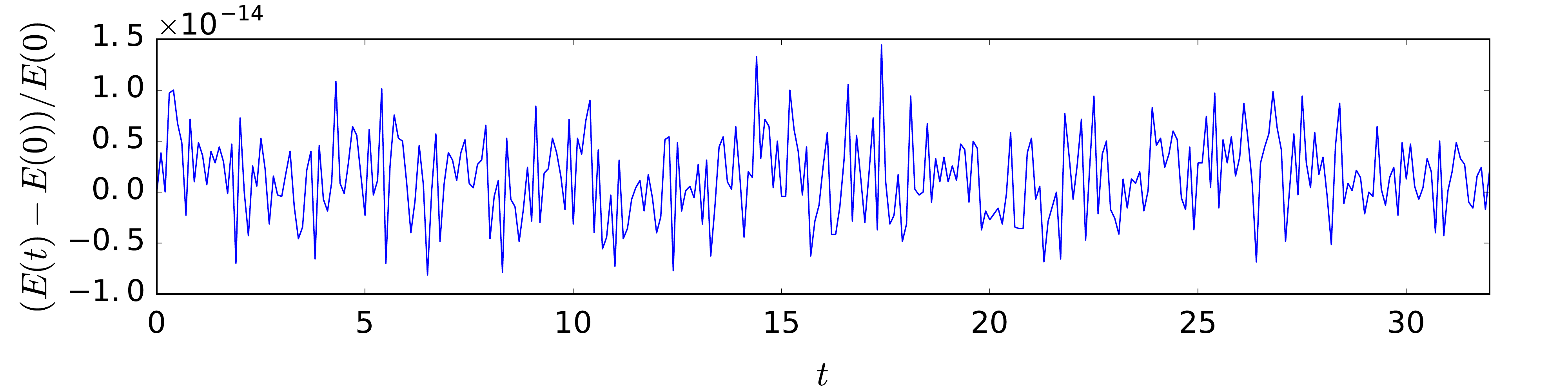}
\includegraphics[width=.45\textwidth]{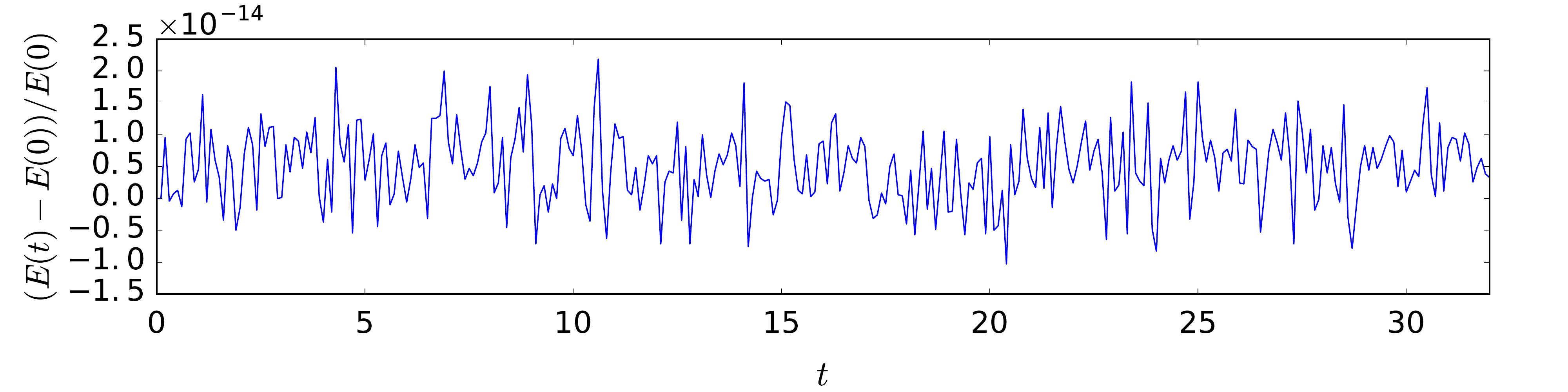}

\includegraphics[width=.45\textwidth]{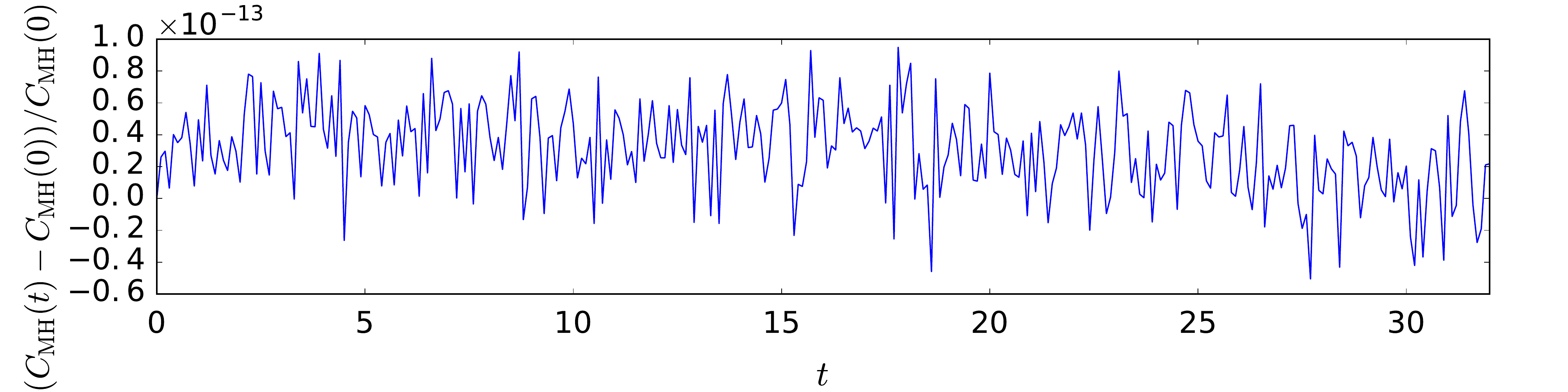}
\includegraphics[width=.45\textwidth]{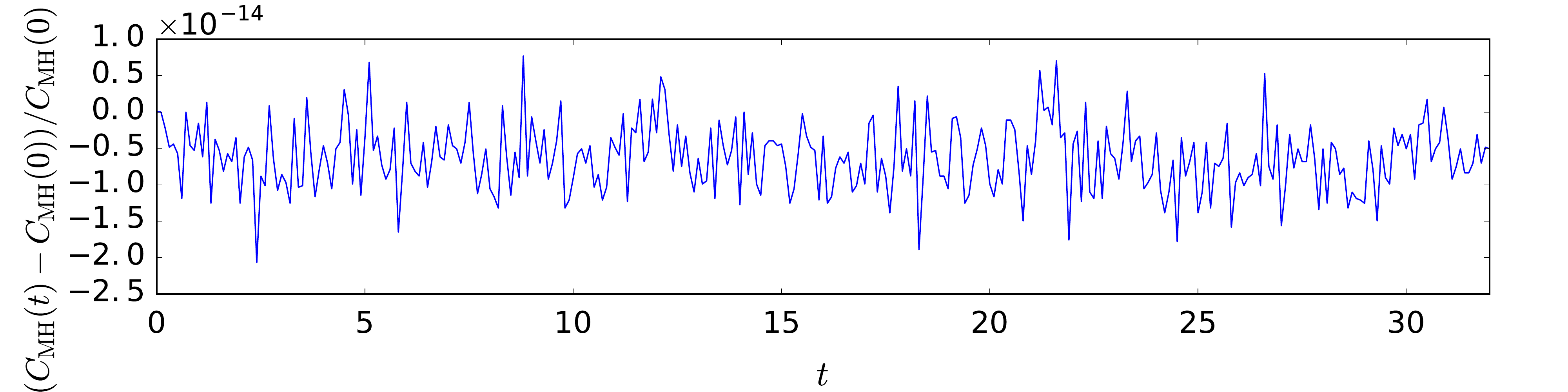}

\includegraphics[width=.45\textwidth]{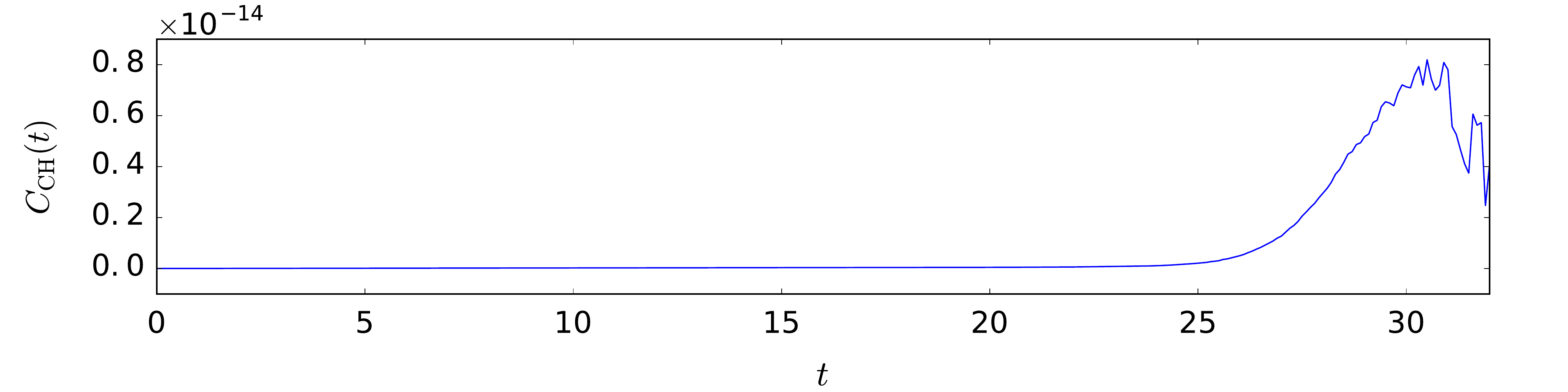}
\includegraphics[width=.45\textwidth]{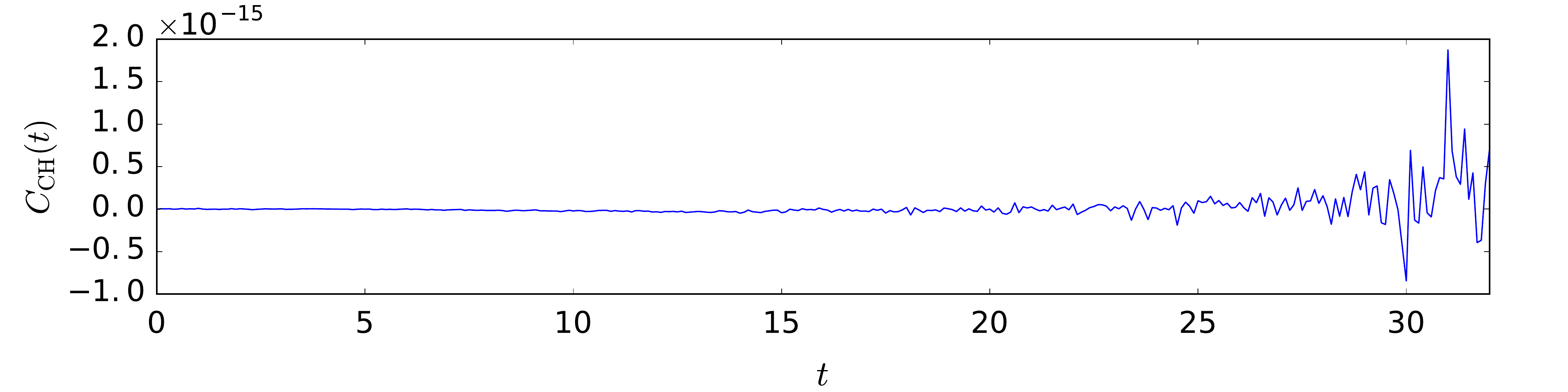}

\caption{Magnetic reconnection with ideal (left) and reduced (right) MHD variational integrator. Error of the total energy $E$, magnetic helicity $C_{\mathrm{MH}}$ and cross helicity $C_{\mathrm{CH}}$.}
\label{fig:mhd_reconnection_timetraces}
\end{figure}

The dynamics of the island growth is practically identically modelled by both variational integrators (see Figure~\ref{fig:mhd_reconnection_psi}). This does not appear too surprising, given the good conservation properties of both methods.
Only after $t=30$ we observe minor differences in the solutions at the inside of the island.
At this point, a secondary Kelvin-Helmholtz-type of instability starts to evolve and the simulations are under-resolved for both integrators.
This instability is followed by a turbulent regime, where energy is continuously transferred to ever finer scales with the constraint of total energy conservation due to the Hamiltonian nature of the system~\cite{Del03,Del05}.
In this situation, from Figure~\ref{fig:mhd_reconnection_psi} it emerges that the variational integrator for reduced MHD does not preserve the intrinsic parity symmetry of the equations, $A(x,y) = A(-x,y)$ and $A(x,y) = A(x,-y)$, anymore. While this is expected in the turbulent regime as a consequence of chaotic dynamics, the ideal MHD integrator seems to preserve this symmetry much better.
On the other hand, the reduced MHD integrator seems to retain more detail of the turbulent fine scale structures along the $x=0$ and more prominently the $y=0$ axes, which appear at about $t=30$. Thus the symmetry breaking is also more pronounced.
That aside, the results of both integrators agree very well in the generalised magnetic potential $\obar{A}$ (see Figure~\ref{fig:mhd_reconnection_psie}) as well as the vorticity $\omega$ and the fields $\obar{B}$ and $V$ (not shown here).
Throughout the simulation, both variational integrators show excellent preservation of energy, the generalised magnetic helicity and cross helicity (see Figure~\ref{fig:mhd_reconnection_timetraces}).

\begin{figure}[p]
\centering
\subfloat[128x64]{
\includegraphics[width=.4\textwidth]{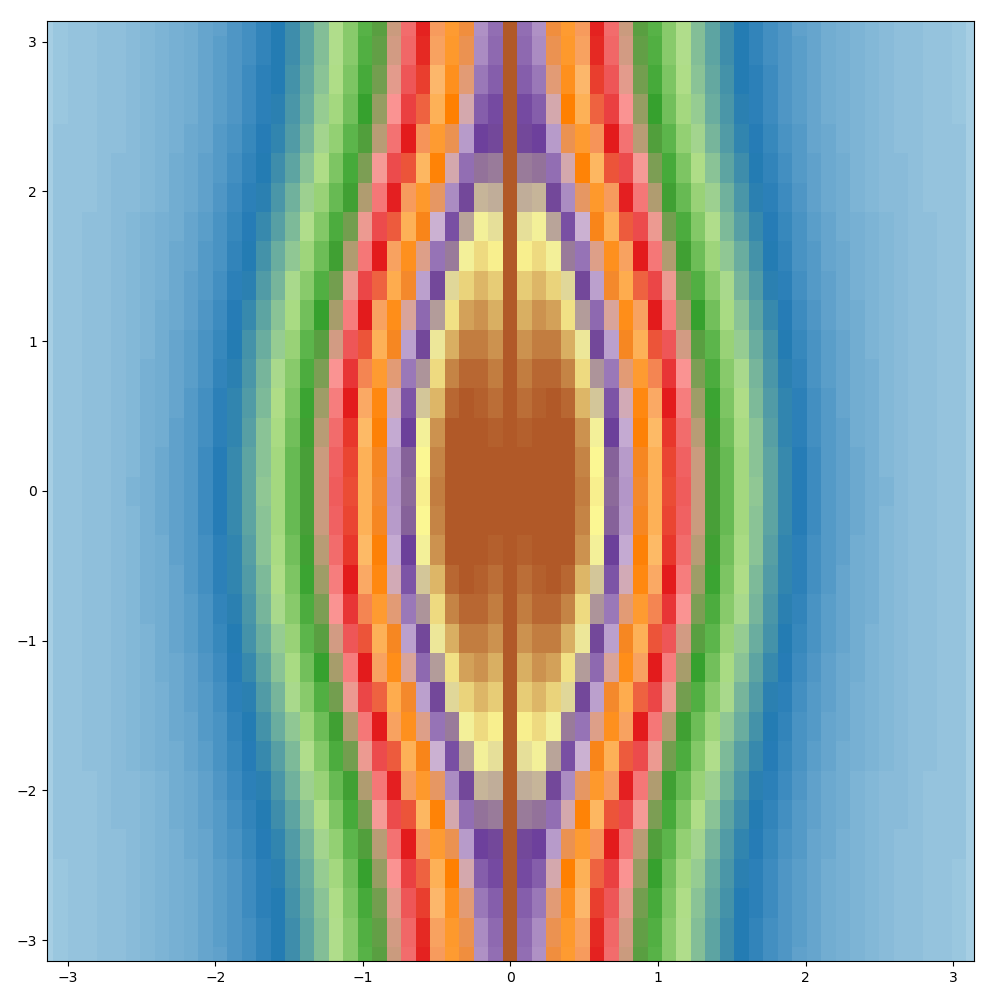}
}
\subfloat[256x128]{
\includegraphics[width=.4\textwidth]{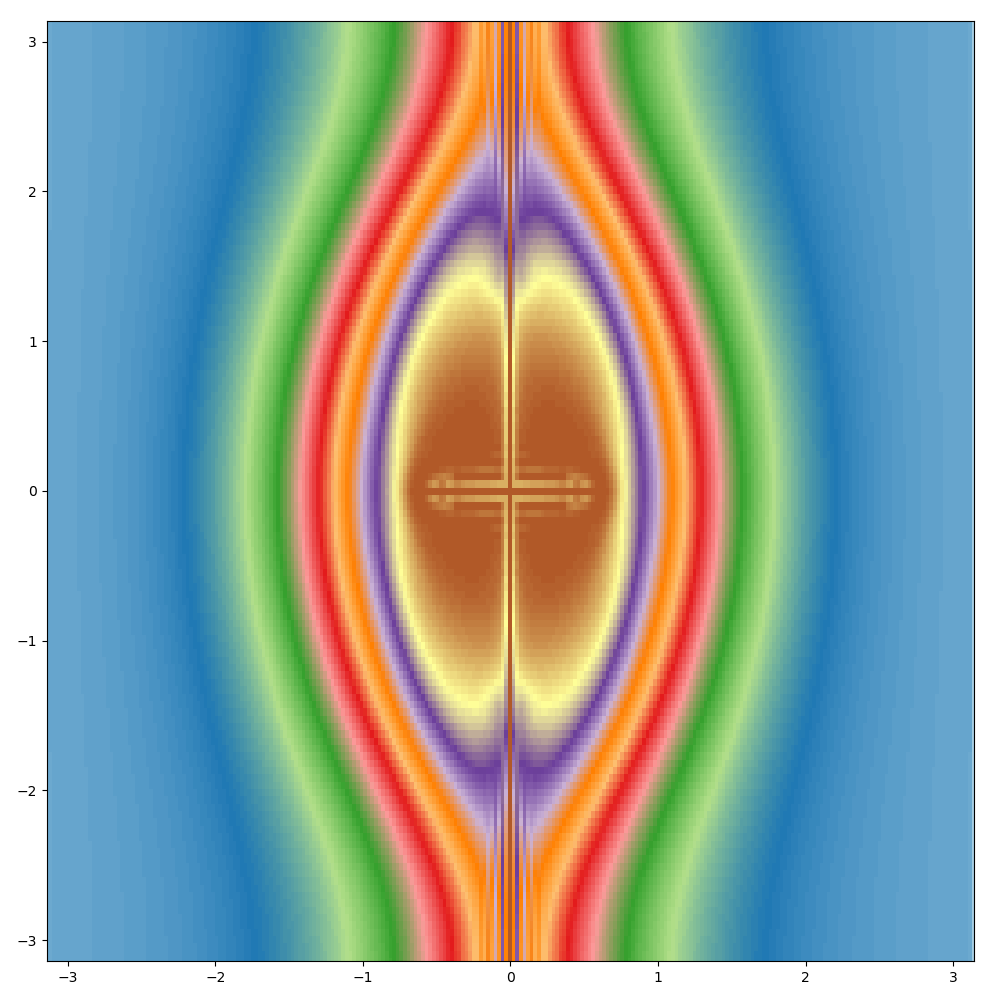}
}

\subfloat[512x256]{
\includegraphics[width=.4\textwidth]{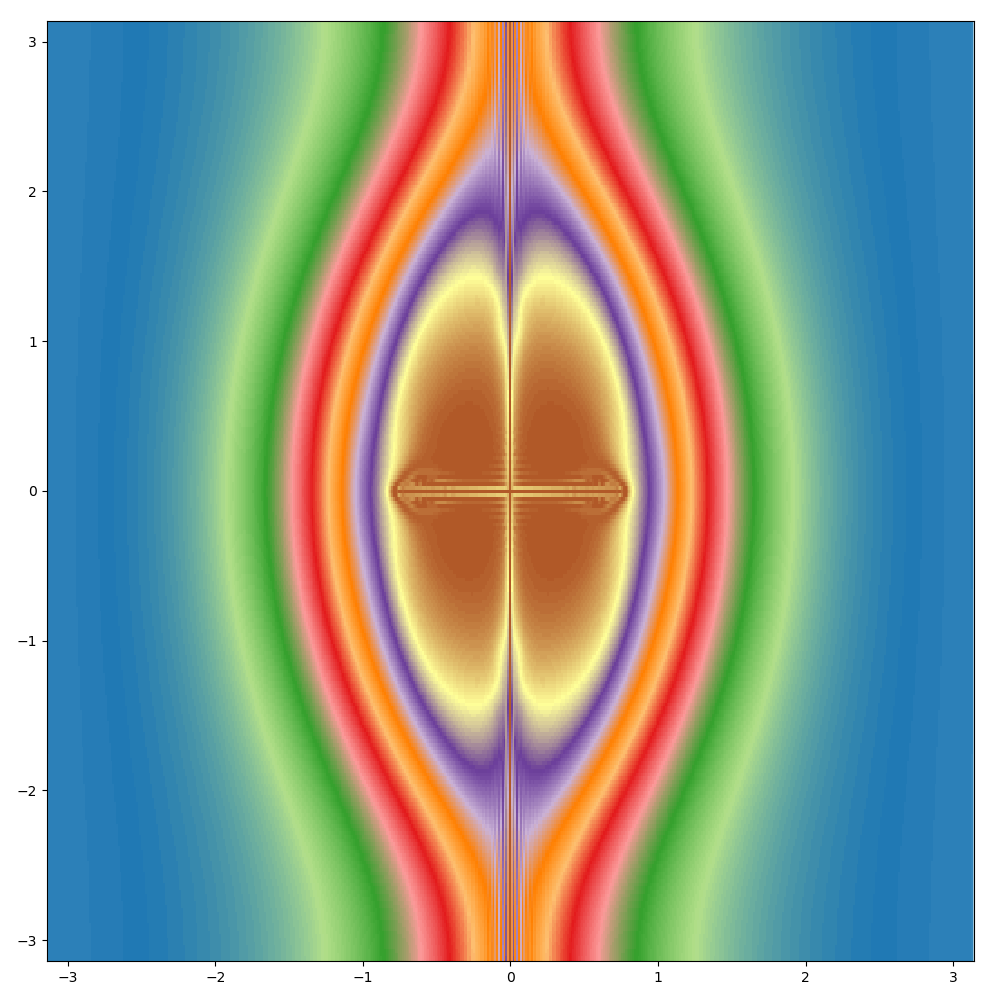}
}
\subfloat[1024x512]{
\includegraphics[width=.4\textwidth]{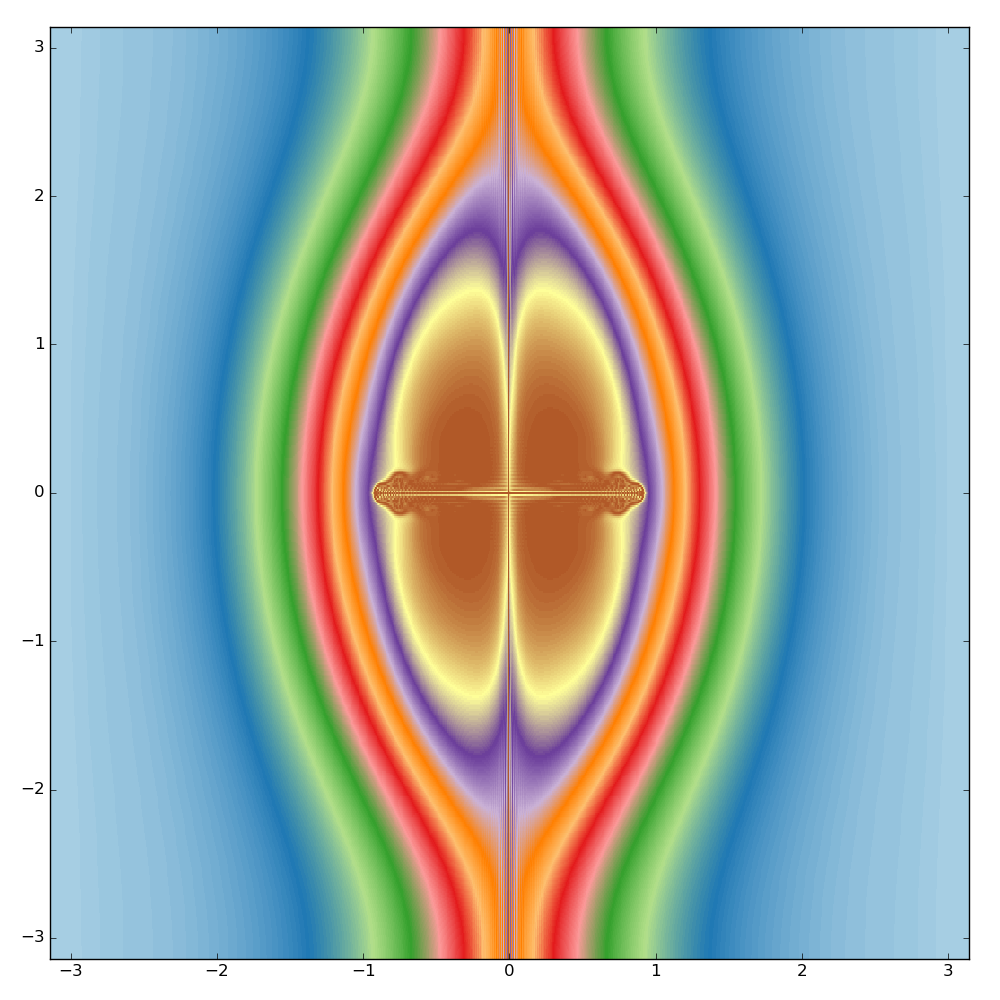}
}

\caption{Magnetic reconnection with ideal MHD variational integrator. Generalised vector potential $\obar{A}$ at $t = 30$ for various resolutions.}
\label{fig:mhd_reconnection_psie_resolution}
\end{figure}

It is worthwhile to note that the variational integrators perform very well even if the simulation is under-resolved. The growth rate of the island is correctly obtained already at a resolution of $256 \times 128$ grid points~(see Figure~\ref{fig:reconnection_growth_rate}). Solely the secondary Kelvin-Helmholtz instability towards the end of the simulation cannot be resolved at lower resolutions.
This stability of the integrators with respect to grid resolution stems from their good conservation properties. In many numerical schemes with dissipation there is no control over the sign of the dissipation, i.e., energy can be spuriously dissipated or fed into the system. The latter usually leads to instabilities or, worse, wrong simulation results. As in the variational integrators such effects are absent, qualitatively and to some extend even quantitatively correct results can be obtained if the simulation is under-resolved (see Figure~\ref{fig:mhd_reconnection_psie_resolution}).

\section{Summary}

In this work we extended the variational integrator for ideal magnetohydrodynamics of~\citet{KrausMaj:2017} to inertial magnetohydrodynamics. To our knowledge, neither an integrator tailored to the structure of this system nor any numerical simulations of the latter have been presented as of yet. 
The formal Lagrangian approach~\cite{KrausMaj:2015} together with a staggered grid, motivated by discrete differential forms, lead to an integrator that guarantees exact conservation of energy, generalised magnetic helicity and cross helicity as well as the divergence of the magnetic field.
These excellent conservation properties have been demonstrated in numerical examples relevant to collisionless reconnection studies.
Particularly remarkable is the absence of artificial magnetic reconnection in the ideal case, even for long integration times. This is expected from the physics, as the magnetic flux is frozen and the magnetic field line topology is preserved, but often this is not what is observed in numerical simulations due to spurious dissipation, especially in simple finite difference schemes like the one presented.
We showed that with the variational integrators reconnection indeed takes place only when electron inertia effects are added.

A theoretical limitation of the presented method is the finite-difference staggered grid approach, which is not easily generalised to higher-order methods and to more complicated geometries.
This work, however, should rather be understood as a proof-of-principle of the applicability of the formal Lagrangian approach~\cite{KrausMaj:2015} to extended magnetohydrodynamics models like inertial MHD.
It is quite remarkable that exact conservation properties can be achieved even with very low-order discretisations and with very low resolutions. The robustness of the integrators with respect to resolution is particularly noteworthy.

For more involved applications, it is straight-forward to apply the ideas presented here in conjunction with other numerical frameworks, such as finite element exterior calculus~\cite{Arnold:2006, Arnold:2010, Christiansen:2011}, mimetic spectral elements~\cite{Gerritsma:2012, Kreeft:2011, Palha:2014} or spline differential forms~\cite{Buffa:2010, Buffa:2011, Ratnani:2012}, in order to obtain numerical schemes of arbitrary order as well as on general meshes.

\section*{Acknowledgements}

Useful discussions with Omar Maj are gratefully acknowledged.
The author has received funding from the European Union's Horizon 2020 research and innovation programme under the Marie Sklodowska-Curie grant agreement No 708124. The views and opinions expressed herein do not necessarily reflect those of the European Commission.

\phantomsection
\addcontentsline{toc}{section}{References}
\bibliographystyle{plainnat}
\bibliography{vi_inertial_mhd2d}

\end{document}